# Energy-Efficient Power Control: A Look at 5G Wireless Technologies

Alessio Zappone, *Member, IEEE*, Luca Sanguinetti, *Senior Member, IEEE*, Giacomo Bacci, *Member, IEEE*, Eduard Jorswieck, *Senior Member, IEEE*, and Mérouane Debbah, *Fellow, IEEE*

***Abstract*—This work develops power control algorithms for energy efficiency (EE) maximization (measured in bit/Joule) in wireless networks. Unlike previous related works, minimum-rate constraints are imposed and the signal-to-interference-plus-noise ratio takes a more general expression, which allows one to encompass some of the most promising 5G candidate technologies. Both network-centric and user-centric EE maximizations are considered. In the network-centric scenario, the maximization of the global EE and the minimum EE of the network are performed. Unlike previous contributions, we develop centralized algorithms that are guaranteed to converge, with affordable computational complexity, to a Karush-Kuhn-Tucker point of the considered non-convex optimization problems. Moreover, closed-form feasibility conditions are derived. In the user-centric scenario, game theory is used to study the equilibria of the network and to derive convergent power control algorithms, which can be implemented in a fully decentralized fashion. Both scenarios above are studied under the assumption that single or multiple resource blocks are employed for data transmission. Numerical results assess the performance of the proposed solutions, analyzing the impact of minimum-rate constraints, and comparing the network-centric and user-centric approaches.**

## I. Introduction

Currently, the percentage of the global world $CO_2$ emissions due to the information and communications technology (ICT) is estimated to be 5% [1]. While this may seem a small percentage, it is rapidly increasing, and the situation will escalate in the near future with the advent of 5G networks. It is anticipated that the number of connected devices will reach 50 billions by 2020 [2], and that a 1000x data rate increase is required to serve so many connected devices [3]. However, it is also clear that obtaining the required 1000x by simply scaling up the transmit powers is not possible, as it would result in an unmanageable energy demand, and in greenhouse gas emissions and electromagnetic pollution above safety thresholds. Instead, the data rate must be increased by a factor 1000, at a similar power consumption as in present networks. This requires a $1000\times$ increase of the energy efficiency (EE), i.e., the efficiency with which ICT systems use energy to transmit data [4]. This is of paramount importance for operators (e.g., to save on electricity bills) and end-users (e.g., to prolong the lifetime of batteries) and thus has motivated a great interest in studying and designing power control strategies taking into account the cost of energy.

Power control for energy efficiency can be performed in a centralized or decentralized manner. Both approaches are of interest in the context of 5G networks, for which network-centric techniques like Cloud-RAN and cooperative multi-point (CoMP) [5], as well as user-centric techniques like device-to-device (D2D) [6] and the use of femto-cells [7], have been proposed. In a network-centric approach, resource allocation is performed centrally and all network-nodes cooperate to maximize a common system-wide performance function. Instead, a user-centric approach is implemented in a distributed way, with the different nodes behaving in a self-organizing, and often competitive, fashion. Network-centric approaches typically grant better performance, with respect to self-organizing algorithms, but on the other hand they are more complex and require a larger feedback overhead to be implemented. An overview of recent results in these two directions is provided below.

### A. State-of-the-art

A non-exhaustive list of recent works dealing with EE optimization in centralized networks includes [8]–[15] and references therein. Given the non-convex, fractional nature of EE, the main mathematical tool for centralized optimization of EE-related metrics is fractional programming – a branch of optimization theory that provides algorithms with polynomial complexity to globally maximize fractional functions with a concave numerator and a convex denominator [16]. However,

A. Zappone and E. Jorswieck are with TU Dresden, Faculty of Electrical and Computer Engineering, Communications Laboratory, Dresden, Germany (email: {alessio.zappone,eduard.jorswieck}@tu-dresden.de).

L. Sanguinetti is with the University of Pisa, Dipartimento di Ingegneria dell'Informazione, Pisa, Italy (luca.sanguinetti@iet.unipi.it) and also with Large Systems and Networks Group (LANEAS), CentraleSupélec, Université Paris-Saclay, 3 rue Joliot-Curie, 91192 Gif-sur-Yvette, France.

G. Bacci is with the Mediterranean Broadband Infrastructure (MBI) srl, Pisa, Italy (email: gbacci@mbigroup.it).

M. Debbah is with Large Systems and Networks Group (LANEAS), CentraleSupélec, Université Paris-Saclay, 3 rue Joliot-Curie, 91192 Gif-sur-Yvette, France and also with the Mathematical and Algorithmic Sciences Lab, Huawei France R&D, Paris, France (e-mail: merouane.debbah@huawei.com).

The work of A. Zappone and E. Jorswieck is funded by the German Research Foundation (DFG), within the projects CEMRIN - grant ZA 747/1-3, and the Collaborative Research Center 912 HAEC, respectively. This research is also supported by the FP7 NEWCOM# project (GA no. 318306). L. Sanguinetti is also funded by the People Programme (Marie Curie Actions) FP7 PIEF-GA-2012-330731 Dense4Green. The research of M. Debbah have been also supported by the ERC Starting Grant 305123 MORE.

Part of this work has been presented at the IEEE International Conference on Communications (ICC15), London, UK, June 8 – 12, 2015.

even this powerful tool fails when interference-limited networks must be optimized. This is due to the fact that the presence of multi-user interference makes the numerator of EE non-concave. A common way out to this problem is to rely on orthogonal or semi-orthogonal transmission schemes as well as on interference cancellation techniques (to fall back to the noise-limited case). Contributions in this direction are given in [9], [14], [15]. In [9], [15] multi-carrier networks are considered, and the global energy efficiency (GEE) of the system (defined as the ratio between the sum achievable rate and the total consumed power) is optimized using orthogonal or semi-orthogonal subcarrier allocation schemes. In [14], the authors consider a multiple-antenna system and aim to maximize the GEE when non-linear interference cancellation techniques are used. However, orthogonal interference suppression schemes inevitably result in a poor resource reuse factor and are not practical in large networks. An alternative approach consists in handling the interference by means of heuristic solutions, typically based on the use of alternating optimization techniques. Examples in this context are given by [10] in which the minimum of the individual EEs is maximized and also by [11], [12] where both the maximization of GEE and of the sum of the individual EEs are considered. While these approaches can operate in interference-limited networks, they do not guarantee convergence and/or are not supported by strong optimality claims. Moreover, they are typically tailored to the maximization of specific EE metrics. A first attempt to provide a unified framework to tackle EE optimization problems in a centralized way is given in [13], where polynomial-time algorithms to optimize the GEE as well as the sum and product of the individual EEs are provided.

As for EE maximization through decentralized solutions, in [17] the authors study the Nash equilibrium problem for a group of players aiming at maximizing their own EE while satisfying power constraints in single and multi-carrier systems, similarly to what was done in [18] for rate maximization. A quasi-variational inequality approach is taken in [19], where power control algorithms for networks with heterogenous users are developed. In [20], [21] a similar problem is considered, with regard to relay-assisted systems, whereas single-user multiple-input multiple-output (MIMO) systems are considered in [22]. However, all of these previous works do not account for rate requirements, and so the resulting users' rates at the equilibrium could be fairly low. Incorporating target rates changes the setting drastically since any user's admissible power allocation policy depends crucially on the policies of all other users. First results in this context are provided in [23] wherein Nash equilibria are found to be the fixed points of a water-filling best-response operator whose water level depends on the rate constraint and circuit power.

### B. Motivation

The aim of this paper is to develop a unified framework for the analysis and design of both centralized (network-centric) and decentralized (user-centric) EE power allocation policies in a wireless network in which $K$ transmitters (possibly) share $N$ mutually orthogonal resource blocks for data transmission. Unlike most previous related works, we aim at maximizing different EE metrics while satisfying minimum rate constraints (or quality-of-service (QoS) requirements). Moreover, we assume that the signal-to-interference-plus-noise-ratio (SINR) experienced by transmitter $k$ at its intended receiver on resource block $n$ takes the following general form:

$$\gamma_{k,n} = \frac{\alpha_{k,n} p_{k,n}}{\sigma^2_{k,n} + \phi_{k,n} p_{k,n} + \sum_{j \neq k} \omega_{kj,n} p_{j,n}} \tag{1}$$

where $p_{k,n}$ is the $k$-th user's transmit power over resource block $n$, whereas $\alpha_{k,n}$, $\phi_{k,n}$, $\omega_{kj,n}$ are positive quantities that do not depend on the users' transmit powers, but only on system parameters and propagation channels. In particular, $\alpha_{k,n}$ and $\phi_{k,n}$ are assumed to depend only on user $k$'s channels on resource block $n$, while the coefficients $\omega_{kj,n}$ depend on the other users' channels on resource block $n$. The main motivation behind the adoption of (1) is that there exist several communication systems and technologies in which the SINR takes this form.[1] Interestingly, this is the case of some candidate technologies for 5G networks, e.g., practical massive MIMO networks in which the massive amount of deployed circuitry prevents the use of high-quality hardware and thus gives rise to hardware impairments [24]. The form in (1) arises also when imperfect channel state information (CSI) is available due to channel estimation errors. This is again a typical situation in practical massive MIMO systems [25]. Other relevant examples are heterogeneous, relay-assisted interference networks (e.g. multi-cell and/or small-cell relay-assisted MIMO networks [26], multi-cell and/or small-cell orthogonal frequency division multiple access (OFDMA) networks [20]), and relay-assisted D2D networks [27]. Finally, other well-established communication technologies are also included, such as ultra wide-band systems [28], or, generally, transmissions affected by inter-symbol interference and frequency-selective fading [29]. In Section II-C we will describe in detail some examples of communication systems using candidate 5G wireless technologies, in which the SINRs take the form in (1).

### C. Contributions and paper outline

The major contributions of this work are as follows:

- A unified framework for EE optimization is developed for both centralized and decentralized networks with rate and power constraints in which the users' SINRs take the more general expression in (1). This allows encompassing some of the emerging technologies for 5G.
- The maximization of the GEE as well as of the minimum EE is considered in the network-centric case. Both problems are non-convex and thus hard to solve. We first derive closed-form feasibility conditions, and then exploit the tools of fractional programming and sequential convex optimization to develop centralized power control algorithms that are guaranteed to converge to a Karush-Kuhn-Tucker (KKT) point of the non-convex problems with affordable computational complexity.

[1] Observe that (1) includes as a special case the SINR expression typically encountered in wireless communication systems, which can be obtained by simply letting $\phi_{k,n} = 0$ for all $k, n$.

- In the decentralized setting, the users in the network are modeled as rational, self-organizing agents that engage in a non-cooperative game wherein each one aims at maximizing its individual EE while targeting its own power and rate constraint. The existence and uniqueness of Nash equilibrium points are studied and a fully distributed algorithm based on best-response dynamics is proposed to reach equilibrium.
- The above scenarios are studied under the assumption that one or more resource blocks are employed for data transmission.

The remainder of this paper is organized as follows. Section II introduces the signal model and formulates the EE maximization problems from both network- and user-centric perspectives. Moreover, Section II-C provides a detailed description of some 5G candidate technologies for which the SINRs are expressed as in (1). The centralized and decentralized approaches for EE maximization in single-resource block transmissions are analyzed in Sections III and IV, respectively. The counterpart cases of multiple-resource blocks are studied in Sections V and VI. In Section VII the performance of the proposed algorithms is numerically analyzed with reference to case-studies inspired to 5G technologies. Concluding remarks are given in Section VIII.

## II. Energy-Efficiency Problem Formulations

Consider a wireless interference network with $K$ transmitters, $S$ receivers, and $N$ available resource blocks (that might represent time or frequency bins) of bandwidth $B$ in which the SINR of user $k$ at its intended receiver takes the general form in (1). The EE $\eta_k$ (measured in bit/Joule) of user $k$ is defined as the ratio of the achievable rate over the $N$ resource blocks and the total consumed power [16]

$$\eta_k \triangleq \frac{\sum_{n=1}^{N} B\log_2(1+\gamma_{k,n})}{p_{c,k}+\mathbf{1}^T\mathbf{p}_k} \tag{2}$$

with $p_{c,k}$ being the circuit power dissipated to operate the $k$-th transmitter and its intended receiver, and $\mathbf{p}_k = [p_{k,1}, p_{k,2}, \ldots, p_{k,N}]^T \in \mathbb{R}_+^N$ being the power allocation vector of user $k$ over the $N$ resource blocks. We assume that $\mathbf{p}_k$ must satisfy the following (local) power constraint:

$$\mathbf{1}^T\mathbf{p}_k - \overline{p}_k \le 0 \tag{3}$$

where $\overline{p}_k$ denotes user $k$'s maximum power. Unlike most previous related works, we assume that minimum achievable rates need to be satisfied. This amounts to setting:

$$\sum_{n=1}^{N}\log_2(1+\gamma_{k,n}) - \underline{\theta}_k \ge 0 \tag{4}$$

where $\underline{\theta}_k$ is the target rate of user $k$ (in bit/s/Hz/user). The feasible set of $\mathbf{p}_k$ is thus given by:

$$\mathcal{P}_k \triangleq \left\{ \mathbf{p}_k \in \mathbb{R}_+^N : \mathbf{1}^T\mathbf{p}_k \le \overline{p}_k, \sum_{n=1}^{N}\log_2(1+\gamma_{k,n}) \ge \underline{\theta}_k \right\}. \tag{5}$$

Accordingly, we call $\boldsymbol{\mathcal{P}} \triangleq \prod_{k=1}^{K}\mathcal{P}_k$ the feasible set of $\mathbf{p} = [\mathbf{p}_1^T, \mathbf{p}_2^T, \ldots, \mathbf{p}_K^T]^T \in \mathbb{R}_+^{KN}$.

### A. Network-centric formulation

Based on the user-centric EE metric (2), two relevant network-centric performance metrics are investigated in this work. The network GEE $\psi$ is given by the system achievable sum-rate over the total power consumed in the system:

$$\psi \triangleq \frac{\sum_{k=1}^{K}\sum_{n=1}^{N} B\log_2(1+\gamma_{k,n})}{p_c + \sum_{k=1}^{K}\mathbf{1}^T\mathbf{p}_k} \tag{6}$$

with $p_c = \sum_{k=1}^{K} p_{c,k}$ being the total circuit power dissipated in the network.[2] Another important energy-efficient metric is the minimum of the weighted EEs, defined as

$$\underline{\eta} \triangleq \min_{k=1,\ldots,K} w_k\eta_k. \tag{7}$$

Within the above setting, the GEE maximization problem can be mathematically formulated as:

$$\psi^\star = \max_{\mathbf{p}\in\boldsymbol{\mathcal{P}}} \quad \psi = \max_{\mathbf{p}\in\boldsymbol{\mathcal{P}}} \quad \frac{\sum_{k=1}^{K}\sum_{n=1}^{N} B\log_2(1+\gamma_{k,n})}{p_c + \sum_{k=1}^{K}\mathbf{1}^T\mathbf{p}_k} \tag{8}$$

whereas the weighted minimum-EE maximization problem can be written as:

$$\underline{\eta}^\star = \max_{\mathbf{p}\in\boldsymbol{\mathcal{P}}} \quad \underline{\eta} = \max_{\mathbf{p}\in\boldsymbol{\mathcal{P}}} \quad \min_{k=1,\ldots,K} w_k\eta_k\ , \tag{9}$$

with $\{w_k\}_k$ non-negative weights. As discussed later, both (8) and (9) are non-convex fractional problems, which will be tackled by means of fractional programming theory[3] and sequential convex optimization.

**Remark 1.** *Observe that $\psi$ and $\underline{\eta}$ represent the two extreme points in the trade-off between global performance and fairness. In particular, $\psi$ can be seen as the benefit-cost ratio of the system, being defined as the ratio between the sum achievable rate and the total consumed power. However, it does not directly depend on the users' EEs, and therefore it does not allow one to tune the individual EEs according to users' needs, as it might be useful in heterogeneous networks. On the other hand, maximizing the (weighted) minimum of the EEs allows us to achieve a fairer resource allocation policy. In particular, it is known that maximizing* (7) *yields a Pareto-efficient point where each quantity $w_k\eta_k$ is the same for all $k$. The whole Pareto-boundary can be simply achieved by varying the weights $\{w_k\}$. However, this usually comes at the price of a performance loss in terms of benefit-cost ratio of the system.*

### B. User-centric formulation

A game-theoretic approach will be taken to solve the user-centric power allocation problem formulated as:

$$\max_{\mathbf{p}_k\in\mathcal{P}_k} \quad \eta_k = \max_{\mathbf{p}_k\in\mathcal{P}_k} \quad \frac{\sum_{n=1}^{N} B\log_2(1+\gamma_{k,n})}{p_{c,k}+\mathbf{1}^T\mathbf{p}_k} \quad \forall k. \tag{10}$$

[2] A slightly more general definition of the GEE considers the weighted sum-rate at the numerator. This might be useful to control the users' individual rates. All the following results can be straightforwardly applied to this definition of GEE.

[3] For completeness, a brief background on fractional programming is provided in Appendix A.

Table I
LIST OF FUNCTIONS AND VARIABLES

| | | | |
|---|---|---|---|
| $\eta_k$ | EE of user $k$ | $\overline{\gamma}_{k,n}$ | SINR in absence of interference and noise |
| $\psi$ | GEE | $\underline{\theta}_k$ | Target rate of user $k$ |
| $\underline{\eta}$ | Minimum-EE | $\underline{\gamma}_k$ | Minimum SINR requirement for user k |
| $\gamma_{k,n}$ | SINR of user $k$ over block $n$ | $\overline{p}_k$ | Maximum power of user $k$ |
| $\mu_{k,n}$ | Equivalent channel gain of user $k$ over block $n$ | $p_c$, $p_{c,k}$ | System and per-user hardware-dissipated power |

For later convenience, we denote by

$$\mu_{k,n} \triangleq \frac{\alpha_{k,n}}{\sigma_{k,n}^2 + \sum\limits_{j\neq k} \omega_{kj,n} p_{j,n}} \tag{11}$$

the equivalent channel gain when $\phi_{k,n} = 0$, and call

$$\overline{\gamma}_{k,n} \triangleq \frac{\alpha_{k,n}}{\phi_{k,n}} \tag{12}$$

the SINR in the absence of interference and thermal noise (i.e., the maximum achievable SINR). Using the above definitions, (1) can be rewritten as[4]

$$\gamma_{k,n} = \frac{\overline{\gamma}_{k,n}\mu_{k,n}p_{k,n}}{\overline{\gamma}_{k,n} + \mu_{k,n}p_{k,n}} \tag{13}$$

or, equivalently,

$$p_{k,n} = \frac{\gamma_{k,n}}{\mu_{k,n}}\left(1 - \frac{\gamma_{k,n}}{\overline{\gamma}_{k,n}}\right)^{-1}. \tag{14}$$

Note that $\gamma_{k,n}$ is a strictly increasing function of $p_{k,n}$ as it easily follows observing that

$$\frac{\partial \gamma_{k,n}}{\partial p_{k,n}} = \frac{\overline{\gamma}_{k,n}^2 \mu_{k,n}}{\left(\overline{\gamma}_{k,n} + \mu_{k,n}p_{k,n}\right)^2} \geq 0. \tag{15}$$

### *C. Applications to 5G technologies*

As mentioned in the introduction, there exist several examples of communication technologies in which the SINR may take the form in (1). Two case-studies are briefly detailed in the sequel.

*1) Massive MIMO:* Consider the uplink (similar results can be obtained for the downlink) of a massive MIMO system composed of $S$ cells wherein the base station (BS) of each cell uses $M$ antennas to communicate with $K$ single-antenna user equipments (UEs). Each UE is associated to a specific serving BS while interfering with all other UEs. As such, a double index notation is used to refer to each UE as e.g., "user $k$ in cell $j$". Under this convention, let us define $\mathbf{h}_{ilj} \in \mathbb{C}^M$ as the channel from UE $j$ in cell $l$ to BS $i$. Denoting by $\mathbf{c}_{ik} \in \mathbb{C}^M$ the receive combining vector of UE $k$ at its intended BS $i$, a lower bound (obtained using a standard bound based on the worst-case uncorrelated noise) of the uplink SINR of UE $k$ in cell $i$ takes the form [30]

$$\gamma_k = \frac{p_{ik}\left|\mathrm{E}\{\mathbf{c}_{ik}^H\mathbf{h}_{iik}\}\right|^2}{p_{ik}\mathrm{var}\{\mathbf{c}_{ik}^H\mathbf{h}_{iik}\} + \mathbf{i}_{ik}} \tag{16}$$

with

$$\mathbf{i}_{ik} = \sigma^2\mathrm{E}\{||\mathbf{c}_{ik}||^2\} + \sum_{(l,j)\neq(i,k)} p_{lj}\mathrm{E}\{|\mathbf{c}_{ik}^H\mathbf{h}_{ilj}|^2\}. \tag{17}$$

Assume that a maximum ratio combining (MRC) receiver is employed for data recovery. This amounts to setting $\mathbf{c}_{ik} = \hat{\mathbf{h}}_{iik}$ where $\hat{\mathbf{h}}_{iik}$ denotes the estimate of $\mathbf{h}_{iik}$ given by

$$\mathbf{h}_{iik} = \hat{\mathbf{h}}_{iik} + \tilde{\mathbf{h}}_{iik} \tag{18}$$

with $\tilde{\mathbf{h}}_{iik}$ being the estimation error statistically independent of $\hat{\mathbf{h}}_{iik}$. Assume that $\mathbf{h}_{ilj} \sim \mathcal{CN}(0, d_{ilj}\mathbf{I}_M)$ where $d_{ilj}$ accounts for the corresponding large-scale channel fading and pathloss from UE $j$ in cell $l$ to BS $i$. If a minimum mean square error (MMSE)-based channel estimation scheme is used at the BS (with full pilot reuse) [30], then we have that $\hat{\mathbf{h}}_{iik} \sim \mathcal{CN}(\mathbf{0}, \rho_{iik}\mathbf{I}_M)$ and $\tilde{\mathbf{h}}_{iik} \sim \mathcal{CN}(\mathbf{0}, (d_{iik} - \rho_{iik})\mathbf{I}_M)$ where

$$\rho_{iik} = \frac{d_{iik}}{\tau + \sum_l d_{ilk}} \tag{19}$$

with $\tau$ being a given parameter that depends on the pilot transmit power and the pilot sequence length. Under the above assumptions, we have that

$$\gamma_k = \frac{p_{ik}\alpha_k}{\sigma_k^2 + p_{ik}\phi_k + \sum\limits_{(l,j)\neq(i,k)} p_{lj}\omega_{kj}} \tag{20}$$

with

$$\alpha_k = \rho_{iik}^2\ , \quad \omega_{kj} = d_{ilj}\rho_{iik} \tag{21}$$

$$\phi_k = d_{iik}\rho_{iik} + \sum\nolimits_{l\neq i} \rho_{ilk}^2 \tag{22}$$

and $\sigma_k^2 = \sigma^2\rho_{iik}$. Except for the cell index (which is only needed to ease understanding), the above SINR turns out to be in the same form of (1), with non-zero coefficients $\phi_k$.

Similar results can be obtained when the system is affected by hardware impairments [24], [25]; for example, unavoidable clock drifts in local oscillators, finite-precision digital-to-analog converters, amplifier non-linearities, finite-order analog filters, and so forth. For the sake of simplicity, let us assume that the hardware impairments are only at the UEs. Following [25], the hardware impairments result in a reduction of the uplink signals by a factor $1 - \epsilon^2$ with $\epsilon$ being the error vector magnitude, and in an additive Gaussian distortion noise which carries the removed useful power. In these circumstances, a lower bound of the achievable SINR can be computed as

$$\gamma_k = \frac{p_{ik}(1-\epsilon^2)\left|\mathrm{E}\{\mathbf{c}_{ik}^H\mathbf{h}_{iik}\}\right|^2}{p_{ik}(1-\epsilon^2)\mathrm{var}\{\mathbf{c}_{ik}^H\mathbf{h}_{iik}\} + p_{ik}\epsilon^2\mathrm{E}\{|\mathbf{c}_{ik}^H\mathbf{h}_{iik}|^2\} + \mathbf{i}_{ik}} \tag{23}$$

[4]Note that (as expected) when there is no self-interference (SI) (i.e., $\phi_{k,n} = 0\ \forall k, n$) $\gamma_{k,n} \to \mu_{k,n}p_{k,n}$.

with $\mathbf{i}_{ik}$ given in (17). Plugging $\mathbf{c}_{ik} = \hat{\mathbf{h}}_{iik}$ into the above equation and taking into account that in the presence of hardware impairments $\hat{\mathbf{h}}_{iik} \sim \mathcal{CN}(\mathbf{0}, \sqrt{1-\epsilon^2}\rho_{iik}\mathbf{I}_M)$ and $\tilde{\mathbf{h}}_{iik} \sim \mathcal{CN}(\mathbf{0}, (d_{iik} - \sqrt{1-\epsilon^2}\rho_{iik})\mathbf{I}_M)$, the SINR is found to be in the same form of (20). The above analysis can also be extended (as shown in the numerical results) to the case in which hardware imperfections are experienced at the BS.

**Remark 2.** *An alternative approach for evaluating the SINR in massive MIMO systems relies on exploiting the large dimensions of the network. Basically, results from random matrix theory are used to compute an asymptotic expression (known as deterministic equivalent) for* (16) *in the limit of* $M, K \to \infty$ *with* $\frac{K}{M} \in (0, 1)$. *Such a deterministic equivalent turns out to be in the same form of* (1) *and close to the effective SINR even for a finite system [24], [30].*

*2) Relay-assisted CoMP interference network:* Consider the uplink of a two-hop multi-point to multi-point network with $S$ BSs each equipped with $M$ antennas and using $N$ subcarriers to communicate with $K$ single-antenna UEs exploiting a single-antenna amplify-and-forward (AF) relay. Denoting by $h_{k,n}^{(r)}$ the channel from user $k$ to the relay on subcarrier $n$, the received signal at the relay can be written as

$$x_{k,n}^{(r)} = \sqrt{p_{k,n}} h_{k,n}^{(r)} b_{k,n} + \sum_{j \neq k} \sqrt{p_{j,n}} h_{j,n}^{(r)} b_{j,n} + \zeta_n^{(r)} \tag{24}$$

where $b_{k,n}$ is the information symbol transmitted by UE $k$ on subcarrier $n$ and $\zeta_n^{(r)} \sim \mathcal{CN}(0, \sigma_n^2)$ is the relay thermal noise. The total power received at the relay on subcarrier $n$ is thus given by

$$\bar{P}_n^{(r)} = \sum_{j=1}^{K} p_{j,n} |h_{j,n}^{(r)}|^2 + \sigma_n^2. \tag{25}$$

In order to avoid amplifier saturation, the received signal needs to be normalized by $\bar{P}_n^{(r)}$ before being amplified by a factor $\sqrt{p_{r,n}}$ and forwarded to the receivers. The signal received at BS $i_k$ associated to transmitter $k$, over subcarrier $n$, is

$$\mathbf{y}_{k,n} = \sqrt{\frac{p_{k,n} p_{r,n}}{\bar{P}_n^{(r)}}} \mathbf{h}_{i_k,n} h_{k,n}^{(r)} b_{k,n} + \mathbf{i}_{k,n} + \mathbf{w}_{i_k,n} \tag{26}$$

where $\mathbf{h}_{i_k,n} \in \mathbb{C}^M$ is the channel vector from the relay to BS $i_k$ on subcarrier $n$ and $\mathbf{i}_{k,n} \in \mathbb{C}^M$ is defined as

$$\mathbf{i}_{k,n} = \sum_{j \neq k} \sqrt{\frac{p_{j,n} p_{r,n}}{\bar{P}_n^{(r)}}} \mathbf{h}_{i_k,n} h_{j,n}^{(r)} b_{j,n} + \sqrt{\frac{p_{r,n}}{\bar{P}_n^{(r)}}} \mathbf{h}_{i_k,n} \zeta_n^{(r)} \tag{27}$$

whereas $\mathbf{w}_{i_k,n} \sim \mathcal{CN}(0, \sigma_{i_k,n}^2 \mathbf{I}_M)$ is the thermal noise at receiver $i_k$ on subcarrier $n$. After linear reception by the filter $\mathbf{c}_{k,n}$ and upon plugging (25) into (26) the SINR takes the form in (1) with

$$\alpha_{k,n} = p_{r,n} |h_{k,n}|^2 |\mathbf{c}_{k,n}^H \mathbf{h}_{i_k,n}|^2 \ , \ \ \phi_{k,n} = \sigma_{i_k,n}^2 |h_{k,n}|^2 \|\mathbf{c}_{k,n}\|^2$$
$$\omega_{kj,n} = \left( p_{r,n} |\mathbf{c}_{k,n}^H \mathbf{h}_{i_k,n}|^2 + \sigma_{i_k,n}^2 \|\mathbf{c}_{k,n}\|^2 \right) |h_{j,n}|^2$$

and $\sigma_{k,n}^2 = \sigma_n^2 (p_{r,n} |\mathbf{c}_{k,n}^H \mathbf{h}_{i_k,n}|^2 + \sigma_{i_k,n}^2 \|\mathbf{c}_{k,n}\|^2)$. We should stress that the above results apply to any interference network in which the transmitters reach the receivers via an AF relay. For example, the above scenario applies to relay-assisted multi-cell and small-cell networks [26], as well as to relay-assisted D2D networks [27].

**Remark 3.** *Observe that in both scenarios described above, the expression of the receive filters* $\mathbf{c}_{ik}$ *and* $\mathbf{c}_{k,n}$ *impacts the SINR expressions through the coefficients* $\alpha_k$, $\phi_k$, $\omega_{kj}$ *and the equivalent noise power* $\sigma_k^2$. *Therefore, any choice of the receive filters that does not depend on the transmit powers, results in a SINR expression as in* (1). *This means that the developed framework can be applied to both MRC and zero-forcing receivers whereas it does not readily apply to MMSE-based receivers as they depend on the transmit powers of UEs through the covariance matrix of the interference.*

## III. Centralized Power Control in Networks with a Single Resource Block

We start our analysis considering the case of a single resource block, which, for example, models single-carrier systems. When $N = 1$, deeper analytical insights can be gained compared to the case with $N > 1$. In particular, the single-resource block scenario is analytically more tractable and allows one to obtain necessary and sufficient feasibility conditions for the centralized energy-efficient optimization problems. On the other hand, only sufficient feasibility conditions can be obtained in the multi-resource block setting (Section V). This also applies to the distributed scenario in terms of more compact existence and uniqueness conditions of the equilibria (Section IV). Instead, the corresponding condition for the multi-resource block setting will be very cumbersome, and thus more difficult to handle. Moreover, the techniques to be presented in this section carry over to scenarios with $N > 1$, and are preparative for the more involved multiple resource block scenario.

Setting $N = 1$ into (1) and (2) and neglecting the block index, (8) and (9) reduce to:

$$\psi^\star = \max_{\mathbf{p} \in \boldsymbol{\mathcal{P}}} \quad \psi = \max_{\mathbf{p} \in \boldsymbol{\mathcal{P}}} \quad \frac{\sum_{k=1}^{K} B \log_2(1 + \gamma_k)}{p_c + \sum_{k=1}^{K} p_k} \tag{28}$$

and

$$\underline{\eta}^\star = \max_{\mathbf{p} \in \boldsymbol{\mathcal{P}}} \quad \underline{\eta} = \max_{\mathbf{p} \in \boldsymbol{\mathcal{P}}} \quad \min_{k=1,\ldots,K} w_k \frac{B \log_2(1 + \gamma_k)}{p_{c,k} + p_k} \tag{29}$$

wherein $\mathbf{p} = [p_1, p_2, \ldots, p_K]^T \in \mathbb{R}_+^K$,

$$\gamma_k = \frac{\alpha_k p_k}{\sigma_k^2 + \phi_k p_k + \sum_{j \neq k} \omega_{kj} p_j} \tag{30}$$

and $\boldsymbol{\mathcal{P}} \triangleq \prod_{k=1}^{K} \mathcal{P}_k$ with

$$\mathcal{P}_k = \{p_k \in \mathbb{R}_+ : p_k \leq \overline{p}_k, \log_2(1 + \gamma_k) \geq \underline{\theta}_k\}. \tag{31}$$

For later convenience, we define

$$\underline{\gamma}_k \triangleq 2^{\underline{\theta}_k} - 1 \tag{32}$$

the minimum SINR requirement for user $k$. Observe that $\underline{\gamma}_k$ must be such that

$$0 \leq \underline{\gamma}_k \leq \overline{\gamma}_k \quad \forall k. \tag{33}$$

The above condition follows observing that when the noise is negligible (i.e., $\sigma_k^2 \to 0$) and only transmitter $k$ is active then (30) reduces to $\gamma_k = \alpha_k/\phi_k = \overline{\gamma}_k$ and thus the rate constraint $\log_2(1+\gamma_k) \geq \underline{\theta}_k$ can be met only if (33) holds true.

### A. Feasibility

The feasibility of (28) and (29) simply amounts to verifying that for given values of $\{\alpha_k\}_k$, $\{\phi_k\}_k$, and $\{w_{k,j}\}_{k\neq j}$, the feasible set $\boldsymbol{\mathcal{P}}$ is not empty. Closed-form necessary and sufficient conditions for $\boldsymbol{\mathcal{P}}$ to be non-empty are provided in the following result.

**Lemma 1.** *Let $\mathbf{F} \in \mathbb{C}^{K\times K}$ be a matrix whose $(k,j)$-th element is defined as*

$$[\mathbf{F}]_{k,j} \triangleq \begin{cases} 0 & j = k \\ \frac{\omega_{kj}\underline{\gamma}_k}{\alpha_k - \phi_k\underline{\gamma}_k} & j \neq k \end{cases} \tag{34}$$

*and denote by $\rho_{\mathbf{F}}$ its spectral radius. The solutions to (28) and (29) exist if and only if*

$$\rho_{\mathbf{F}} < 1 \quad \text{and} \quad (\mathbf{I}-\mathbf{F})^{-1}\underline{\mathbf{s}} \leq \overline{\mathbf{p}} \tag{35}$$

*where $\overline{\mathbf{p}} = [\overline{p}_1, \overline{p}_2, \ldots, \overline{p}_K]^T \in \mathbb{R}_+^{K\times 1}$ and $\mathbf{s} \in \mathbb{R}_+^K$ has elements given by $[\underline{\mathbf{s}}]_k \triangleq \sigma_k^2\underline{\gamma}_k(\alpha_k - \phi_k\underline{\gamma}_k)^{-1}$.*

*Proof:* Following the same steps of the proof of Lemma 2 in [31] allows us to prove the sufficiency of (35). For the necessity, assume that there is a vector $\mathbf{p}'$ satisfying the target rates but such that $p'_j \geq \overline{p}_j$ for some $j$. Since $\gamma_k$ is a stricly increasing function of $p_k$, we have that

$$\frac{\alpha_k p_k}{\sigma_k^2 + \phi_k p_k + \sum_{j\neq k}\omega_{kj}p'_j} < \overline{\gamma}_k \tag{36}$$

for any $p_k \leq \overline{p}_k$. Otherwise stated, there exists no power $p_k \leq \overline{p}_k$ such that $\gamma_k = \overline{\gamma}_k$ . Now, suppose that $\rho_{\mathbf{F}} \geq 1$. Since $\mathbf{F}$ is non-negative, from [32, Theorem 2.1] it follows that there does not exist a power vector $\mathbf{p} \geq \mathbf{0}$ such that $\gamma_k = \overline{\gamma}_k$. This proves that (35) is also necessary. ■

### B. GEE maximization

As described in Appendix A, fractional programming provides efficient tools to maximize ratios in which the numerator is a concave function, the denominator is a convex function, and the constraint set is convex, whereas no low-complexity optimization method is available if any of these properties is not met. Unfortunately, the objective function in (28) does not have a concave numerator, and therefore finding the global solution of (28) with affordable complexity appears difficult. To overcome this difficulty, we integrate fractional programming theory with the framework of sequential convex programming [33]. This allows us to develop a computationally-efficient algorithm which is guaranteed to converge to a first-order optimal solution of (28). The general idea of sequential convex programming is to find local optima of a difficult problem with objective $f$ to maximize, by solving a sequence of easier problems with objectives $\{f_i\}_i$. In the generic $i$-th step of the sequence, we require the following three properties:

1) $f_i(\mathbf{x}) \leq f(\mathbf{x})$, for all $\mathbf{x}$;
2) $f_i(\mathbf{x}^{(i-1)}) = f(\mathbf{x}^{(i-1)})$;
3) $\nabla f_i(\mathbf{x}^{(i-1)}) = \nabla f(\mathbf{x}^{(i-1)})$.

wherein $\mathbf{x}^{(i-1)}$ denotes the maximizer of $f_{i-1}$. This approach has been used for resource allocation in wireless networks in [34], [35] for rate maximization, and, more recently, in [13], [26], [36] for EE maximization in different settings.

The critical issue of this approach is to find suitable approximations $\{f_i\}_i$ which fulfill the listed requirements, while at the same time resulting in simpler optimization problems. As far as GEE maximization is concerned, this can be accomplished by leveraging the following lower-bound of the logarithmic function [37]. Specifically, $\forall\gamma, \tilde{\gamma} \geq 0$ we have that

$$\log_2(1+\gamma) \geq a\log_2\gamma + b \tag{37}$$

with

$$a = \frac{\tilde{\gamma}}{1+\tilde{\gamma}} \quad b = \log_2(1+\tilde{\gamma}) - \frac{\tilde{\gamma}}{1+\tilde{\gamma}}\log_2\tilde{\gamma}. \tag{38}$$

The right-hand side (RHS) and left-hand side (LHS) of (37) are equal at $\gamma = \tilde{\gamma}$, and the same holds for their derivatives with respect to $\gamma$ evaluated at $\gamma = \tilde{\gamma}$. Therefore, we may lower-bound $\psi$ as follows:

$$\psi \geq \frac{\sum_{k=1}^{K} B\left[a_k\log_2(\gamma_k) + b_k\right]}{p_c + \sum_{k=1}^{K} p_k} = \frac{\sum_{k=1}^{K} B\left[b_k + a_k\log_2(\alpha_k p_k)\right]}{p_c + \sum_{k=1}^{K} p_k} - \frac{\sum_{k=1}^{K} B\left[a_k\log_2\left(\sigma_k^2 + \phi_k p_k + \sum_{j\neq k}\omega_{kj}p_j\right)\right]}{p_c + \sum_{k=1}^{K} p_k} = \tilde{\psi} \tag{39}$$

from which, letting $p_k = 2^{q_k}$, one gets

$$\tilde{\psi} = \frac{\sum_{k=1}^{K} B\left[b_k + a_k\log_2(\alpha_k) + a_k q_k\right]}{p_c + \sum_{k=1}^{K} 2^{q_k}} - \frac{\sum_{k=1}^{K} B\left[a_k\log_2\left(\sigma_k^2 + \phi_k 2^{q_k} + \sum_{j\neq k}\omega_{kj}2^{q_j}\right)\right]}{p_c + \sum_{k=1}^{K} 2^{q_k}}. \tag{40}$$

Using the above results, the solution to (28) can be lower bounded as

$$\psi^\star \geq \tilde{\psi}^\star = \max_{\mathbf{q}\in\boldsymbol{\mathcal{Q}}} \quad \tilde{\psi} \tag{41}$$

with $\mathbf{q} = [q_1, q_2, \ldots, q_K]^T$, $\boldsymbol{\mathcal{Q}} = \prod_{k=1}^{K}\mathcal{Q}_k$ and

$$\mathcal{Q}_k = \{q_k \in \mathbb{R} : 2^{q_k} \leq \overline{p}_k, \log_2(1+\gamma_k) \geq \underline{\theta}_k\}. \tag{42}$$

Observe now that for any given $\{a_k\}_k$ and $\{b_k\}_k$, the numerator and denominator of (40) are both differentiable, and respectively concave[5] and convex in $\{q_k\}_k$. Finally, the set $\mathcal{Q}_k$ can be shown to be convex for all $k$. Indeed, the $k$-th rate constraint can be equivalently rewritten as

$$2^{q_k}(\alpha_k - \underline{\gamma}_k\phi_k) \geq \underline{\gamma}_k\left(\sigma_k^2 + \sum_{j\neq k}\omega_{kj}2^{q_j}\right). \tag{43}$$

[5] Recall that the log-sum-exp function is convex [38].

**Algorithm 1** Network-centric EE maximization for $N = 1$

Test feasibility by Lemma 1.
**if** Feasible **then**
  Set $i = 0$ and choose any $\mathbf{p}^{(0)} \in \boldsymbol{\mathcal{P}}$;
  Set $\tilde{\gamma}_k^{(0)} = \gamma_k^{(0)}(\mathbf{p}^{(0)})$ and compute $a_k^{(0)}$, $b_k^{(0)}$ as in (38);
  **repeat**
    $i = i + 1$;
    **if** GEE **then**
      Solve (41) with parameters $a_k^{(i-1)}$ and $b_k^{(i-1)}$ and set $\{q_k^{(i)}\}_k = \arg\max \tilde{\psi}_i$, $p_k^{(i)} = 2^{q_k^{(i)}}$;
    **end if**
    **if** Minimum EE **then**
      Solve (45) with parameters $a_k^{(i)}$ and $b_k^{(i)}$ and set $\{q_k^{(i)}\}_k = \arg\max \underline{\tilde{\eta}}_i$, $p_k^{(i)} = 2^{q_k^{(i)}}$;
    **end if**
    Set $\tilde{\gamma}_k^{(i)} = \gamma_k(\mathbf{p}^{(i)})$ and compute $a_k^{(i)}$, $b_k^{(i)}$ as in (38);
  **until** convergence
**end if**

Since (33) must hold true, one gets (applying the logarithm function to both sides)

$$q_k - \log_2\left(\sigma_k^2 + \sum_{j \neq k} \omega_{kj} 2^{q_j}\right) + \log_2\left(\frac{\alpha_k - \underline{\gamma}_k \phi_k}{\underline{\gamma}_k}\right) \geq 0 \quad (44)$$

which turns out to be a convex constraint. As a consequence, (41) is a fractional problem which can be globally and efficiently solved by means of fractional programming tools [16], such as the Dinkelbach's algorithm [39]. This leads to the general iterative procedure formulated in Algorithm 1 whose convergence is proved in Appendix B.

**Proposition 1.** *Algorithm 1 monotonically increases the GEE value and converges to a point fulfilling the KKT conditions of the original problem* (28).

### *C. Weighted Minimum EE Maximization*

The key difference between (28) and (29) is that the objective function $\underline{\eta}$ in (29) involves $K$ fractional functions $\{\eta_k\}$ rather than a single one. This makes (29) fall within the framework of *generalized* fractional programming, which studies the maximization of functions of multiple ratios. In this more general scenario, Dinkelbach's algorithm fails, even assuming that each ratio $\{\eta_k\}$ has a concave numerator and a convex denominator. Instead, the problem can be tackled using an extension of Dinkelbach's algorithm known as Generalized Dinkelbach's algorithm (see Appendix A), which is guaranteed to converge to the global solution of a max-min fractional problem with limited complexity, provided each ratio has a concave numerator and a convex denominator [40]. Next, we show how the generalized Dinkelbach's procedure together with sequential convex optimization can be successfully applied to solve (29).

To begin with, observe that the $\min(\cdot)$ function is increasing so that the inequality in (37) can be used to lower-bound the solution to (29) as

$$\underline{\eta}^\star \geq \underline{\tilde{\eta}}^\star = \max_{\mathbf{q} \in \boldsymbol{\mathcal{Q}}} \min_{k=1,\ldots,K} w_k \tilde{\eta}_k \quad (45)$$

where

$$\tilde{\eta}_k = B\frac{\left[b_k + a_k \log_2(\alpha_k) + a_k q_k\right]}{p_{c,k} + 2^{q_k}} - B\frac{\left[a_k \log_2\left(\sigma_k^2 + \phi_k 2^{q_k} + \sum_{j \neq k} \omega_{kj} 2^{q_j}\right)\right]}{p_{c,k} + 2^{q_k}} \quad (46)$$

and $q_k$ is still given by $q_k = \log_2 p_k$. Since each ratio in (46) has a concave numerator and a convex denominator, $\underline{\tilde{\eta}}^\star$ can be computed by means of the Generalized Dinkelbach's algorithm, and the maximization of $\underline{\eta}$ can be tackled as in Algorithm 1, whose convergence is stated in the following proposition and proved in Appendix C.

**Proposition 2.** *Algorithm 1 monotonically increases the value of $\underline{\eta}$ and converges to a point fulfilling the KKT conditions of the epigraph-form representation of the original problem* (29).

**Remark 4.** *Algorithm 1 can be straightforwardly specialized to maximize the system sum rate and the minimum of the users' rates, since these two metrics coincide with the numerator of the GEE and with the minimum of the numerators of the users' EEs, respectively.*

## IV. Distributed Power Control in Networks with a Single Resource Block

A decentralized power control algorithm looks for the solution of the following coupled problems [17], [20], [23]:

$$\arg \max_{p_k \in \mathcal{P}_k(\mathbf{p}_{-k})} \eta_k(p_k, \mathbf{p}_{-k}) \quad \forall k \quad (47)$$

where $\mathbf{p}_{-k} = [p_1, \ldots, p_{k-1}, p_{k+1}, \ldots, p_K]^T$ is the interference vector containing all powers except user $k$'s, and $\mathcal{P}_k(\mathbf{p}_{-k})$ is defined as in (31). This problem can be formulated as the non-cooperative game in normal form:

$$\mathcal{G} \triangleq \{\mathcal{K}, \{\mathcal{P}_k(\mathbf{p}_{-k})\}_k, \{\eta_k(p_k, \mathbf{p}_{-k})\}_k\} \quad (48)$$

where (in game theory parlance) $\mathcal{K} = [1, 2, \ldots, K]$ is the set of players, $\mathcal{P}_k(\mathbf{p}_{-k})$ is player $k$'s strategy set, $u_k(\mathbf{p}) = \eta_k(p_k, \mathbf{p}_{-k})$ is player $k$'s utility function. The $K$ coupled problems in (47) define the best-response dynamics (BRD) of the game, while the solution of the $k$-th problem in (47) is the $k$-th player's best-response to the other players' choices. More formally, let us define the best response $\mathcal{B}_k(\mathbf{p}_{-k})$ of player $k$ to an interference vector $\mathbf{p}_{-k}$ (or, equivalently, $\mu_k$ as easily follows from (11)) as

$$\mathcal{B}_k(\mathbf{p}_{-k}) \triangleq \arg \max_{p_k \in \mathcal{P}_k(\mathbf{p}_{-k})} \eta_k(p_k, \mathbf{p}_{-k}). \quad (49)$$

Any fixed point of the BRD is a Nash equilibrium of the game. In general a non-cooperative game might admit zero, one, or more equilibria, and even if one or more equilibria exist, the convergence of the BRD is not guaranteed. As a consequence, crucial issues in the analysis of a non-cooperative game are to establish the existence and uniqueness of an equilibrium, and whether implementing the BRD eventually yields an equilibrium. In our scenario, answering these questions is more challenging due to the fact that, unlike regular non-cooperative games, not only the utility functions, but also the players' strategy sets are mutually coupled, depending on the other players' actions $\mathbf{p}_{-k}$. A similar non-cooperative game is termed a *generalized* non-cooperative game, and more restrictive conditions have to be fulfilled for a (unique) generalized Nash equilibrium (GNE) to exist and for the BRD to converge. To begin with, the following result is given:

**Lemma 2.** *If*

$$\overline{p}_k \geq \underline{\gamma}_k \frac{\sigma_k^2 + \sum_{j\neq k}\omega_{kj}\overline{p}_j}{\alpha_k - \phi_k \underline{\gamma}_k} \quad \forall k \tag{50}$$

*then* $\mathcal{B}_k(\mathbf{p}_{-k})$ *takes the form*

$$\mathcal{B}_k(\mathbf{p}_{-k}) = \min\left\{\overline{p}_k, \max\left\{p_k^\star, \underline{p}_k\right\}\right\} \tag{51}$$

*wherein*

$$\underline{p}_k(\mathbf{p}_{-k}) \triangleq \frac{\underline{\gamma}_k}{\mu_k}\left(1 - \frac{\underline{\gamma}_k}{\overline{\gamma}_k}\right)^{-1} = \underline{\gamma}_k \frac{\sigma_k^2 + \sum_{j\neq k}\omega_{kj}p_j}{\alpha_k - \phi_k\underline{\gamma}_k} \tag{52}$$

*and*

$$p_k^\star \triangleq \arg\max_{p_k\in\mathbb{R}_+} \eta_k\left(p_k, \mathbf{p}_{-k}\right). \tag{53}$$

*Proof:* The first part of the thesis easily follows from rewriting the rate constraints $\gamma_k \geq \underline{\gamma}_k$ (using (30)) as

$$p_k \geq \underline{\gamma}_k \frac{\sigma_k^2 + \sum_{j\neq k}\omega_{kj}p_j}{\alpha_k - \phi_k\underline{\gamma}_k}. \tag{54}$$

Since $p_k \leq \overline{p}_k$ for all $k \in \mathcal{K}$, then

$$\underline{\gamma}_k \frac{\sigma_k^2 + \sum_{j\neq k}\omega_{kj}\overline{p}_j}{\alpha_k - \phi_k\underline{\gamma}_k} \geq \underline{\gamma}_k \frac{\sigma_k^2 + \sum_{j\neq k}\omega_{kj}p_j}{\alpha_k - \phi_k\underline{\gamma}_k}. \tag{55}$$

Hence, if $\forall k \in \mathcal{K}$ (50) holds, then there always exists a power $p_k \in [0, \overline{p}_k]$ such that $\gamma_k \geq \underline{\gamma}_k$ is fulfilled. The last part of the proof follows by leveraging [20], where it is shown that for any given $\mathbf{p}_{-k}$, $\eta_k$ is unimodal and thus admits a unique maximizer $p_k \in \mathbb{R}_+$. Accounting for the power and rate constraints and imposing (50) eventually yields (51). ■

### A. Analysis of the Equilibria

The existence and uniqueness of the GNE points of $\mathcal{G}$ are now studied under the assumption that (50) holds.

**Proposition 3.** *The game* $\mathcal{G}$ *admits a nonempty set of GNE points.*

*Proof:* Observe that the existence of a GNE is guaranteed under the following assumptions [41]:

1) The players' feasible action sets $\mathcal{P}_k(\mathbf{p}_{-k})$ are nonempty, closed, convex, and contained in some compact set $\mathcal{C}_k$ for all $\mathbf{p}_{-k} \in \mathcal{P}_{-k} \equiv \prod_{\ell\neq k}\mathcal{P}_\ell$.
2) The sets $\mathcal{P}_k(\mathbf{p}_{-k})$ vary continuously with $\mathbf{p}_{-k}$ (in the sense that the graph of the set-valued correspondence $\mathbf{p}_{-k} \mapsto \mathcal{P}_k(\mathbf{p}_{-k})$ is closed).
3) Each user's payoff function $\eta_k(p_k, \mathbf{p}_{-k})$ is quasi-concave in $p_k$ for all $\mathbf{p}_{-k} \in \mathcal{P}_{-k}$.

In our setting, if the sufficient condition (50) is satisfied, then the sets $\mathcal{P}_k(\mathbf{p}_{-k})$ are nonempty, convex,[6] closed and bounded for every $\mathbf{p}_{-k}$. Moreover, each of them varies continuously with $\mathbf{p}_{-k}$ since the rate constraint $\log_2(1+\gamma_k) \geq \underline{\theta}_k$ in $\mathcal{P}_k(\mathbf{p}_{-k})$ is itself continuous in $\mathbf{p}_{-k}$. Finally, following [20] $\eta_k(p_k, \mathbf{p}_{-k})$ is proved to be strictly pseudo-concave since it is given by the ratio between a strictly concave and an affine function. Since any strictly pseudo-concave function is also quasi-concave [16], the third condition is fulfilled. ■

The following result shows that a unique GNE exists, and that the BRD always converges to such point.

**Proposition 4.** *The game* $\mathcal{G}$ *admits a unique GNE point, which can be obtained by starting from any feasible power vector* $\{p_k\}_{k=1}^K$ *and iteratively updating the transmit powers according to* (51).

*Proof:* The proof builds upon the standard function framework [42], which states that a non-cooperative game admits a unique equilibrium (reachable by iteratively computing the players' best-responses) provided the game admits at least one equilibrium and the best-response function is a standard function.[7] Since we have already shown that the game admits a GNE (see Proposition 3), we are left with proving that (51) is a standard function for all $k$. Towards this end, $p_k^\star(\mathbf{p}_{-k})$ is proved to be standard in [20, Appendix A]. As for the function $\underline{p}_k(\mathbf{p}_{-k})$ in (52), it is non-negative because $\underline{\gamma}_k \leq \overline{\gamma}_k$, and it also fulfills the monotonicity property because it is increasing in all $\{p_j\}_{j\neq k}$. As for the scalability property, take any $\beta > 1$, then it holds

$$\underline{p}_k(\beta\mathbf{p}_{-k}) = \beta\underline{\gamma}_k \frac{\frac{\sigma_k^2}{\beta} + \sum_{j\neq k}\omega_{kj}p_j}{\alpha_k - \phi_k\underline{\gamma}_k} \tag{56}$$

$$< \beta\underline{\gamma}_k \frac{\sigma_k^2 + \sum_{j\neq k}\omega_{kj}p_j}{\alpha_k - \phi_k\underline{\gamma}_k} = \beta\underline{p}_k(\mathbf{p}_{-k}). \tag{57}$$

Finally, since both $p_k^\star(\mathbf{p}_{-k})$ and $\underline{p}_k(\mathbf{p}_{-k})$ are standard functions, and since $\overline{p}_k$ does not depend on $\mathbf{p}_{-k}$, we may conclude that (51) is also a standard function because both $\max(\cdot)$ and $\min(\cdot)$ are increasing functions. ■

### B. Distributed implementation

The best response of a generic player $k$ is characterized in the sequel to come up with an iterative algorithm that allows each player to reach the GNE in a distributed manner. Toward this end, let us first define

$$\nu_k(x) \triangleq \overline{\gamma}_k\left[1 + \frac{x}{2B\mu_k}\left(\overline{\gamma}_k - g_k(x)\right)\right]^+ \tag{58}$$

and

$$g_k(x) \triangleq \sqrt{\overline{\gamma}_k^2 + \frac{4B\mu_k}{x}\left(1+\overline{\gamma}_k\right)} \tag{59}$$

with $\mu_k$ and $\overline{\gamma}_k$ given by (11) and (12), respectively.

[6]Note that the constraint function $\log_2(1+\gamma_k)$ is concave in $p_k$.

[7]Recall that a vector function $g(\mathbf{p})$ is standard if it fulfills the properties of *i)* non-negativity: $g(\mathbf{p}) \geq 0$ for all $\mathbf{p}$; *ii)* monotonicity: $g(\mathbf{p}_1) \geq g(\mathbf{p}_2)$ for all $\mathbf{p}_1 \succeq \mathbf{p}_2$; *iii)* scalability: $g(\beta\mathbf{p}) < \beta g(\mathbf{p})$, for all $\mathbf{p}$ and $\beta > 1$.

**Algorithm 2** Iterative algorithm to solve (47).

**initialize** $i = 0$ and $\forall k$ $p_k[0] \in \mathbb{R}_+$ in the feasible set
**repeat**
  **for** $k = 1$ to $K$ **do**
    **receive** $\gamma_k[i]$ from the serving access point
    **compute** $\mu_k[i]$ using (64)
    **use** $\mu_k[i]$ to update $\underline{p}_k[i]$ in (63)
    **use** $\mu_k[i]$ in (61) to run the Dinkelbach's algorithm
    **set** $\lambda_k^\star[i]$ equal to the Dinkelbach's output and update the power as:

$$p_k[i+1] = \min\left\{\overline{p}_k, \max\left\{\pi_k\left(\lambda_k^\star[i]\right), \underline{p}_k[i]\right\}\right\}$$

  **end for**
  **update** $i = i + 1$
**until** convergence

**Lemma 3.** *For any given $\mathbf{p}_{-k}$ (or, equivalently, $\mu_k$), the solution to* (53) *is found to be*

$$p_k^\star = \pi_k\left(\lambda_k^\star\right) \triangleq \frac{\nu_k\left(\lambda_k^\star\right)}{\mu_k}\left(1 - \frac{\nu_k\left(\lambda_k^\star\right)}{\overline{\gamma}_k}\right)^{-1} \tag{60}$$

*where $\lambda_k^\star$ is obtained through the Dinkelbach's algorithm as the solution of the following equation:*

$$B\log_2\left(1 + \nu_k\left(\lambda_k^\star\right)\right) - \lambda_k^\star\left(p_{c,k} + \pi_k\left(\lambda_k^\star\right)\right) = 0. \tag{61}$$

*Proof:* The proof is given in Appendix D. ∎

Denote by $p_k[i]$ the transmit power of the $k$-th player at the $i$-th iteration step. By virtue of Proposition 4 and Lemma 3, it follows that an iterative algorithm operating according to

$$p_k[i+1] = \min\left\{\overline{p}_k, \max\left\{\pi_k\left(\lambda_k^\star[i]\right), \underline{p}_k[i]\right\}\right\} \tag{62}$$

where $\underline{p}_k[i]$ is computed as (using (52))

$$\underline{p}_k[i] = \frac{\underline{\gamma}_k}{\mu_k[i]}\left(1 - \frac{\underline{\gamma}_k}{\overline{\gamma}_k}\right)^{-1} \tag{63}$$

converges to the unique GNE of $\mathcal{G}$, with $\mu_k[i]$ being the equivalent channel gain in (11) at the $i$-th iteration step. The pseudo-code is reported in Algorithm 2.

A close inspection of (58) – (61) and (63) reveals that the computation of $p_k[i+1]$ through (62) only requires knowledge of $\mu_k[i]$. Although not available at the $k$-th terminal, this information can be easily acquired taking into account that:

$$\mu_k[i] = \frac{\gamma_k[i]}{p_k[i]}\left(1 - \frac{\gamma_k[i]}{\overline{\gamma}_k}\right)^{-1} \tag{64}$$

where $\gamma_k[i]$ denotes the SINR of transmitter $k$ measured at its intended receiver at iteration $i$. Since $p_k[i]$ and $\overline{\gamma}_k$ are locally available at the transmitter, the computation of $\mu_k[i]$ only requires knowledge of $\gamma_k[i]$. The latter can be easily estimated at the receiver and sent back to the corresponding transmitter via a return downlink channel. Therefore, besides being guaranteed to converge to the unique GNE, Algorithm 2 can also be implemented in a fully decentralized fashion.

# V. Centralized power control in networks with multiple resource blocks

In this section, we turn our attention to the case in which each transmitter can use multiple resource blocks, i.e., $N > 1$. Differently from the case in which $N = 1$, the rate constraints in (8) and (9) are not in a convex form. Nevertheless, the methodology used in Section III can be successfully extended to find sufficient feasibility conditions and to derive low complexity algorithms that converge to KKT points.

## A. GEE maximization

Following the same steps of Section III-B, we leverage the lower bound in (37) to obtain

$$\psi \geq \tilde{\psi} = \frac{\sum_{k=1}^{K}\sum_{n=1}^{N} B\left[b_{k,n} + a_{k,n}\log_2\left(\alpha_{k,n}\right) + a_{k,n}q_{k,n}\right]}{p_c + \sum_{k=1}^{K}\sum_{n=1}^{N} 2^{q_{k,n}}} - \frac{\sum_{k=1}^{K}\sum_{n=1}^{N} B\left[a_{k,n}\log_2\left(\sigma_{k,n}^2 + \phi_{k,n}2^{q_{k,n}} + \sum_{j\neq k}\omega_{kj,n}2^{q_{j,n}}\right)\right]}{p_c + \sum_{k=1}^{K}\sum_{n=1}^{N} 2^{q_{k,n}}} \tag{65}$$

with $q_{k,n} = \log_2 p_{k,n}$. Although the numerator and denominator of $\tilde{\psi}$ in (65) are again jointly concave and convex in $\{q_{k,n}\}_{k,n}$ (as in the case of a single resource block), the rate constraints in (8) are not in a convex form yet, due to the sum over the multiple resource blocks. To overcome this issue, we resort to the same trick and lower-bound the LHS of the rate constraint by the concave function,

$$\sum_{n=1}^{N}\log_2(1+\gamma_{k,n}) \geq \sum_{n=1}^{N}\left[b_{k,n} + a_{k,n}\log_2\left(\alpha_{k,n}\right) + a_{k,n}q_{k,n}\right] - \sum_{n=1}^{N}\left[a_{k,n}\log_2\left(\sigma_{k,n}^2 + \phi_{k,n}2^{q_{k,n}} + \sum_{j\neq k}\omega_{kj,n}2^{q_{j,n}}\right)\right] = \tilde{R}_k. \tag{66}$$

Finally, we may write

$$\psi^\star \geq \tilde{\psi}^\star = \max_{\mathbf{q}\in\mathcal{Q}} \quad \tilde{\psi} \tag{67}$$

with $\mathcal{Q} = \prod_{k=1}^{K}\mathcal{Q}_k$ and $\mathcal{Q}_k$ being now given by

$$\mathcal{Q}_k = \left\{\mathbf{q}_k \in \mathbb{R}^N : \sum_{n=1}^{N} 2^{q_{k,n}} \leq \overline{p}_k, \tilde{R}_k \geq \underline{\theta}_k\right\}. \tag{68}$$

The solution to the above problem can be computed by means of Dinkelbach's algorithm, which leads to the power allocation procedure illustrated in Algorithm 3. Using similar arguments as in Proposition 1, the following result can be proved:[8]

**Proposition 5.** *Algorithm 3 monotonically increases the GEE value and converges to a point fulfilling the KKT conditions of the original non-convex problem* (8)*.*

[8]As observed in the proof of Proposition 2, the sequential convex optimization tool allows us to find a KKT point of the original problem also when the constraint functions are lower-bounded together with the objective.

**Algorithm 3** Network-centric EE maximization for $N > 1$

1: Set $i = 0$ and choose a starting point $\mathbf{p}^{(0)}$;
2: For any $k$ and $n$, set $\tilde{\gamma}_{k,n}^{(0)} = \gamma_{k,n}^{(0)}(\mathbf{p}^{(0)})$ and compute $a_{k,n}^{(0)}$ and $b_{k,n}^{(0)}$ as in (38);
3: **if** (67) or (71) is unfeasible **then**
4: **break**
5: **end if**
6: **repeat**
7: $i = i + 1$;
8: **if** GEE **then**
9: Solve (67) with parameters $a_{k,n}^{(i-1)}$ and $b_{k,n}^{(i-1)}$ and set $\{q_{k,n}^{(i)}\} = \arg\max \tilde{\psi}_i$, $p_{k,n}^{(i)} = 2^{q_{k,n}^{(i)}}$;
10: **end if**
11: **if** Minimum EE **then**
12: Solve (71) with parameters $a_{k,n}^{(i)}$ and $b_{k,n}^{(i)}$ and set $\{q_{k,n}^{(i)}\} = \arg\max \tilde{\underline{\eta}}_i$, $p_{k,n}^{(i)} = 2^{q_{k,n}^{(i)}}$;
13: **end if**
14: $\tilde{\gamma}_{k,n}^{(i)} = \gamma_{k,n}(\mathbf{p}^{(i)})$ and compute $a_{k,n}^{(i)}$ and $b_{k,n}^{(i)}$ as in (38);
15: **until** convergence

### *B. Weighted minimum EE maximization*

A similar approach as in Section V-A can be used to solve (9). In particular, exploiting the fact that the $\min(\cdot)$ function is increasing, using (37), and setting $p_{k,n} = 2^{q_{k,n}}$ allows us to lower-bound $\underline{\eta}$ as

$$\underline{\eta} \geq \min_{k=1,\ldots,K} w_k \tilde{\eta}_k = \tilde{\underline{\eta}} \tag{69}$$

where

$$\tilde{\eta}_k = \frac{\sum_{n=1}^{N} B\left[b_{k,n} + a_{k,n}\log_2\left(\alpha_{k,n}\right) + a_{k,n}q_{k,n}\right]}{p_{c,k} + \sum_{n=1}^{N} 2^{q_{k,n}}} - \frac{\sum_{n=1}^{N} B\left[a_{k,n}\log_2\left(\sigma_{k,n}^2 + \phi_{k,n}2^{q_{k,n}} + \sum_{j\neq k}\omega_{kj,n}2^{q_{j,n}}\right)\right]}{p_{c,k} + \sum_{n=1}^{N} 2^{q_{k,n}}}. \tag{70}$$

Then, we have that

$$\underline{\eta}^{\star} \geq \tilde{\underline{\eta}}^{\star} = \max_{\mathbf{q}\in\mathcal{Q}} \quad \min_{k=1,\ldots,K} w_k \tilde{\eta}_k. \tag{71}$$

Observe that $\tilde{\eta}_k$ in (70) has a concave numerator and a convex denominator, meaning that (71) can be globally solved by means of the Generalized Dinkelbach's algorithm. The resulting power allocation procedure is given in Algorithm 3. By a similar reasoning as in Proposition 2, the following result can be proved.

**Proposition 6.** *Algorithm 3 monotonically increases the $\underline{\eta}$ value and converges to a point fulfilling the KKT conditions of the epigraph-form representation of the original non-convex problem* (9).

**Remark 5** (Feasibility of (8) and (9))**.** *Observe that Line 3 in Algorithm 3 is also a sufficient feasibility test for* (8) *and* (9) *since both are guaranteed to be feasible provided* (67) *and* (71) *are feasible, which can be checked by means of a convex feasibility test. Moreover, this implies that* (67) *and* (71) *will remain feasible for all iterations of Algorithm 3. Observe also that similar necessary and sufficient feasibility conditions as in Section III can in principle be derived if per-resource-block QoS constraints are considered. This amounts to enforcing $\log_2(1+\gamma_{k,n}) \geq \theta_{k,n}$ with $\theta_{k,n}$ being such that $\sum_{n=1}^{N}\theta_{k,n} = \theta_k$, for all $k$, with $\{\theta_k\}_k$ given target rates.*

## VI. Distributed Power Control in Networks with Multiple Resource Blocks

As done for the single resource block case, we define

$$\nu_{k,n}(x) \triangleq \frac{\overline{\gamma}_{k,n}}{2\mu_{k,n}}\left[2\mu_{k,n} + \frac{x}{B}\left(\overline{\gamma}_{k,n} - g_{k,n}(x)\right)\right]^{+} \tag{72}$$

and

$$g_{k,n}(x) \triangleq \sqrt{\overline{\gamma}_{k,n}^2 + \frac{4B\mu_{k,n}}{x}\left(1+\overline{\gamma}_{k,n}\right)}. \tag{73}$$

Let us also define $\underline{\mathbf{p}}_k$ as the power vector minimizing the transmit power while satisfying the rate constraints. Mathematically, we have that:

$$\underline{\mathbf{p}}_k \triangleq \arg\max_{\mathbf{p}_k\in\mathbb{R}_+^N} \quad \mathbf{1}^T\mathbf{p}_k \tag{74}$$

$$\text{subject to} \quad \sum_{n=1}^{N}\log_2(1+\gamma_{k,n}) - \underline{\theta}_k \geq 0$$

from which (using the same arguments of Appendix D) one gets:

$$\underline{p}_{k,n} = \pi_{k,n}\left(\underline{\lambda}_k\right) \quad \forall n \tag{75}$$

with $\underline{\lambda}_k$ being such that:

$$\sum_{n=1}^{N}\log_2\left(1+\nu_{k,n}\left(\underline{\lambda}_k\right)\right) - \theta_k = 0 \tag{76}$$

and

$$\pi_{k,n}(x) \triangleq \frac{\nu_{k,n}(x)}{\mu_{k,n}}\left(1 - \frac{\nu_{k,n}(x)}{\overline{\gamma}_{k,n}}\right)^{-1}. \tag{77}$$

**Lemma 4.** *For any given $\mathbf{p}_{-k}$, the entries of $\boldsymbol{\mathcal{B}}_k(\mathbf{p}_{-k}) \in \mathbb{R}_+^N$ are found to be*

$$\left[\boldsymbol{\mathcal{B}}_k(\mathbf{p}_{-k})\right]_n = \min\left\{\overline{p}_{k,n}, \max\left\{\pi_{k,n}\left(\lambda_k^{\star}\right), \underline{p}_{k,n}\right\}\right\} \tag{78}$$

*where $\lambda_k^{\star}$ is obtained through the Dinkelbach method as the solution of the following equation:*

$$\sum_{n=1}^{N} B\log_2\left(1+\nu_{k,n}\left(\lambda_k^{\star}\right)\right) - \lambda_k^{\star}\left(p_{c,k} + \sum_{n=1}^{N}\pi_{k,n}\left(\lambda_k^{\star}\right)\right) = 0. \tag{79}$$

*Proof:* The proof relies on similar arguments (omitted for space limitations) of those in Appendix D for $N = 1$. ■

Observe that (78) can be equivalently rewritten as (since $\pi_{k,n}(\cdot)$ in (77) is a strictly decreasing function):

$$\left[\boldsymbol{\mathcal{B}}_k(\mathbf{p}_{-k})\right]_n = \pi_{k,n}\left(\lambda_k'\right) \tag{80}$$

with

$$\lambda_k' \triangleq \max\left\{\overline{\lambda}_k, \min\left\{\lambda_k^{\star}, \underline{\lambda}_k\right\}\right\} \tag{81}$$

and $\overline{\lambda}_k$ such that $\forall n \; \pi_{k,n}\left(\overline{\lambda}_k\right) = \overline{p}_{k,n}$. A sufficient condition for (80) (or, equivalently, for (78)) to be a contraction mapping

**Algorithm 4** Algorithm to reach the GNE of $\mathcal{G}$ for $N > 1$.

**initialize** $i = 0$ and $\forall k, n$ $p_{k,n}[0] \in \mathbb{R}_+$ in the feasible set
**repeat**
  **for** $k = 1$ to $K$ **do**
    **receive** $\{\gamma_{k,n}[i]\}_n$ from the intended receiver
    **compute** $\{\mu_{k,n}[i]\}_n$ using (84)
    **use** $\{\mu_{k,n}[i]\}_n$ to compute $\underline{\lambda}_k$ in (76) by means of the Dinkelbach algorithm
    **use** $\underline{\lambda}_k$ to update $\{\underline{p}_{k,n}[i]\}_n$ in (75)
    **use** $\{\mu_{k,n}[i]\}_n$ in (79) to run the Dinkelbach algorithm
    **set** $\lambda_k^\star[i]$ equal to the Dinkelbach output and update the power as:

$$p_{k,n}[i+1] = \min\left\{\overline{p}_{k,n}, \max\left\{\pi_{k,n}\left(\lambda_k^\star[i]\right), \underline{p}_{k,n}[i]\right\}\right\}$$

  **end for**
  **update** $i = i + 1$
**until** $\forall k, n$ $p_{k,n}[i] = p_{k,n}[i-1]$

is provided in the following proposition (see Appendix E for the proof), thus ensuring the existence and uniqueness of a GNE for $\mathcal{G}$ when $N > 1$, and the convergence of the BRD.

**Proposition 7.** *The EE game $\mathcal{G}$ admits a unique GNE and the BRD converges to such equilibrium whenever, for all $k$*

$$\sum_{\substack{j=1\\j\neq k}}^{K}\sum_{n=1}^{N}\omega_{kj,n}^2 \sup_{\boldsymbol{\mu}_k\in\boldsymbol{\Omega}_k}\left\{\sum_{\ell,n\in\mathcal{S}_k^\star}\left[s_{k,n}\mathbb{1}_{n=\ell} + \frac{\overline{\gamma}_{k,n}}{g_{k,n}}\xi_{k,\ell}\right]^2\right\} < 1 \quad (82)$$

*where $\boldsymbol{\Omega}_k \triangleq \prod_{n=1}^N (0, \alpha_{k,n}]$ and $\xi_{k,\ell} \triangleq \frac{\mu_{k,\ell}^2}{\alpha_{k,\ell}}\frac{\partial(1/\lambda_k')}{\partial\mu_{k,\ell}}$ whereas $\mathcal{S}_k^\star \triangleq \{n = 1, \ldots, N : \mu_{k,n} > \lambda_k'\}$ and $s_{k,n}$ is defined as*

$$s_{k,n} \triangleq \overline{\gamma}_{k,n}\frac{g_{k,n}\left(\lambda_k'\right) - \left(2 + \overline{\gamma}_{k,n}\right)}{2\alpha_{k,n}\left(1 + \overline{\gamma}_{k,n}\right)} - \frac{\overline{\gamma}_{k,n}\mu_{k,n}}{\lambda_k'\alpha_{k,n}g_{k,n}\left(\lambda_k'\right)}. \quad (83)$$

Similarly to the single resource block case, denote by $p_{k,n}[i]$ the transmit power of the $k$-th player over block $n$ at the $i$-th iteration step and, accordingly, compute $\mu_{k,n}[i]$ as follows:

$$\mu_{k,n}[i] = \frac{\gamma_{k,n}[i]}{p_{k,n}[i]}\left(1 - \frac{\gamma_{k,n}[i]}{\overline{\gamma}_k}\right)^{-1} \quad (84)$$

with $\gamma_{k,n}[i]$ being the SINR of transmitter $k$ over block $n$ measured at its intended receiver at iteration $i$. From the results of Proposition 7, it follows that the iterative procedure illustrated in Algorithm 4 converges to the unique GNE of $\mathcal{G}$, and can be implemented in a fully decentralized fashion.

## VII. Numerical Results

Numerical results are now given to assess the performance of the proposed solutions. To this end, two case-studies are considered, namely, a hardware-impaired massive MIMO system and a multi-carrier relay-assisted interference network.

### A. Hardware-Impaired Massive MIMO System

Consider the uplink of a massive MIMO system as described in Section II-C1, with $K = 5$, $S = 1$, and $M = 50$. The communication bandwidth is $B = 1$ MHz, and MRC is used for data recovery under the assumption of perfect channel estimation. We also set $\epsilon = 1$, thereby considering that no hardware impairments are present at the UEs. However, we assume that hardware impairments are present at the BS, and we denote by $\epsilon_{BS}$ the resulting error magnitude. As observed in Section II-C1, this scenario results again in an SINR as in (1). Indeed, following [43] and by similar steps as in Section II-C1, the $k$-th UE's SINR is written as in (1) with

$$\alpha_k = \left|\mathbf{h}_k^H\mathbf{h}_k\right|^2, \quad \phi_k = \epsilon_{BS}^2\mathbf{h}_k^H\mathbf{D}_k\mathbf{h}_k \quad (85)$$

$$\omega_{kj} = |\mathbf{h}_k^H\mathbf{h}_j|^2 + \epsilon_{BS}^2\mathbf{h}_k^H\mathbf{D}_j\mathbf{h}_k \quad (86)$$

with $\mathbf{D}_j = \mathrm{diag}(\{|h_j(m)|^2\}_{m=1}^M)$ and $\sigma_k^2 = \sigma^2\mathbf{h}_k^H\mathbf{h}_k$. All channels are generated according to the Rayleigh fading model with path-loss model as in [44]. All UEs have the same maximum feasible power $\overline{p}_k = \overline{P}$ $\forall k$ and hardware-dissipated power $p_{c,k} = 10$ dBm $\forall k$. The receive noise power is $\sigma^2 = FB\mathcal{N}_0$, with $F = 3$ dB and $\mathcal{N}_0 = -174$ dBm/Hz. The minimum rate constraint $\underline{\theta}_k$ is set as a percentage $R_k\%$ of the maximum rate that user $k$ can achieve when $p_k \to \infty$, while the other users' powers are finite, namely:

$$\theta_k = \frac{R_k}{100}\log_2(1 + \overline{\gamma}_k) = \frac{R_k}{100}\log_2\left(1 + \frac{\alpha_k}{\phi_k}\right) \quad (87)$$

For simplicity, we assume $R_k\% = R\%$ for all users $k$.

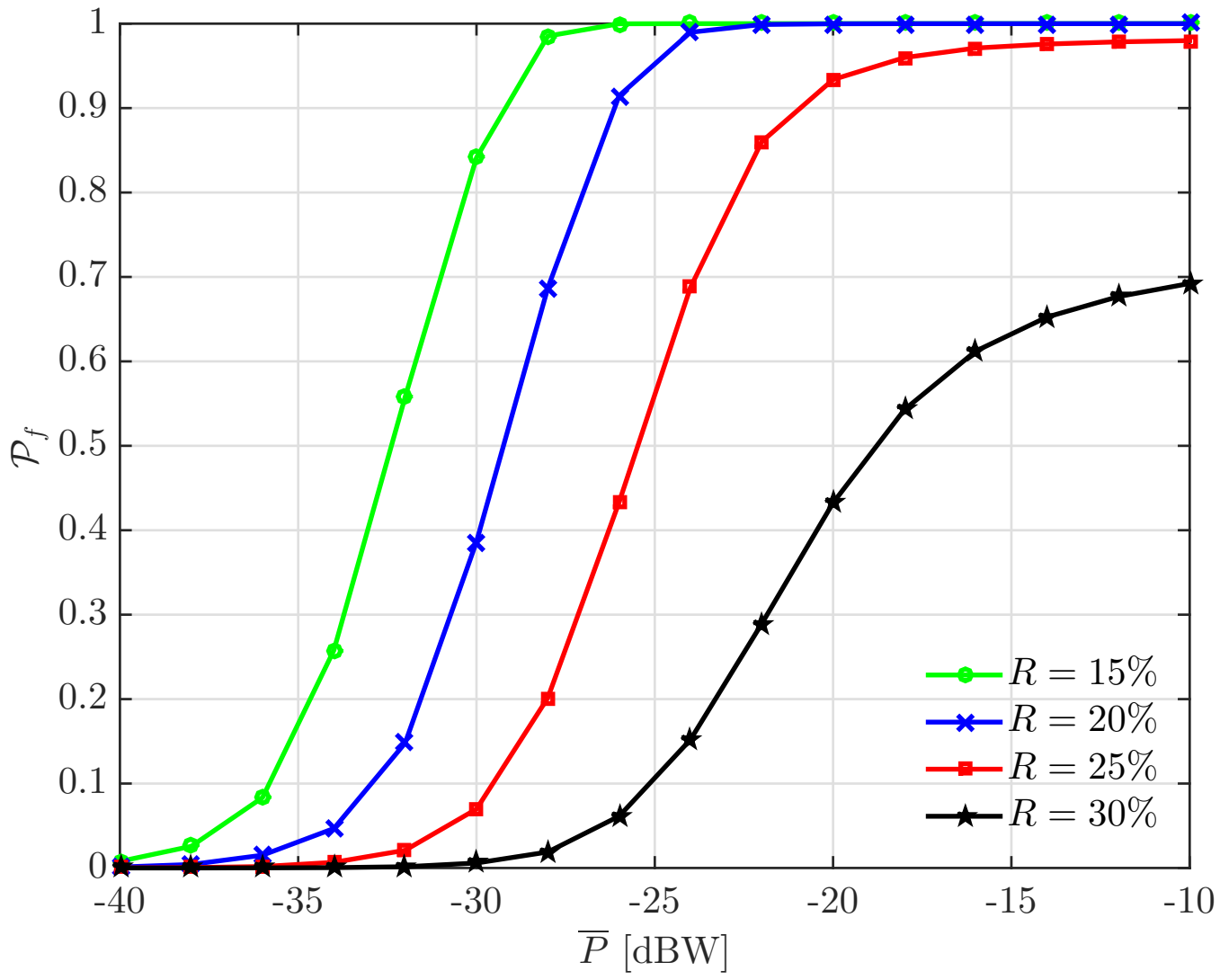

Figure 1. $K = 5$; $M = 50$; $\tau = 10^{-2}$. Probability of feasibility $\mathcal{P}_f$ versus $\overline{P}$ with minimum per user-rate constraints: (a) $R = 15\%$; (b) $R = 20\%$; (c) $R = 25\%$; (d) $R = 30\%$.

We begin by analyzing the feasibility probability $\mathcal{P}_f$ of the EE maximization problems as a function of $\overline{P}$ for different values of $R$. The results are obtained by averaging over $5 \cdot 10^4$ independent scenarios of users' drops and channel coefficients. As seen, $\mathcal{P}_f$ approaches 1 for realistic values of $\overline{P}$ up to 25% of the maximum rate, whereas for $R = 30\%$ the typical transmit power levels which are used in the uplink of present cellular systems, are not enough to ensure a high $\mathcal{P}_f$.

Next, we analyze the system performance in terms of EE and data rate. Fig. 2 shows the average GEE for the following resource allocation policies: (a) Algorithm 1 with $R = 20\%$; (b) Algorithm 1 without QoS constraints, i.e.

Table II
$K=5; M=50; \tau=10^{-2}$. AVERAGE NUMBER OF REQUIRED ITERATIONS TO REACH CONVERGENCE VERSUS $P_{max}$ FOR: (A) ALGORITHM 1 FOR GEE MAXIMIZATION WITH $R=20\%$; (B) ALGORITHM 1 FOR GEE MAXIMIZATION WITH $R=0\%$; (C) ALGORITHM 2 FOR DISTRIBUTED RESOURCE ALLOCATION WITH $R=20\%$; (D) ALGORITHM 2 FOR DISTRIBUTED RESOURCE ALLOCATION WITH $R=0\%$.

| | Maximum power $\overline{P}$ [dBW] | | | | | | | |
|---|---|---|---|---|---|---|---|---|
| | $\overline{P}=-38$ | $\overline{P}=-34$ | $\overline{P}=-30$ | $\overline{P}=-26$ | $\overline{P}=-22$ | $\overline{P}=-18$ | $\overline{P}=-14$ | $\overline{P}=-10$ |
| Algorithm 1. $R=0\%$ | 2.63 | 3.69 | 4.68 | 6.30 | 6.53 | 6.49 | 6.50 | 6.51 |
| Algorithm 1. $R=20\%$ | 2.63 | 3.67 | 4.87 | 6.68 | 6.70 | 6.76 | 6.76 | 6.77 |
| Algorithm 2. $R=0\%$ | 1 | 1.01 | 1.07 | 1.42 | 2.54 | 3.66 | 4.12 | 4.50 |
| Algorithm 2. $R=20\%$ | 1 | 1.01 | 1.07 | 1.42 | 2.54 | 3.67 | 4.19 | 6.71 |

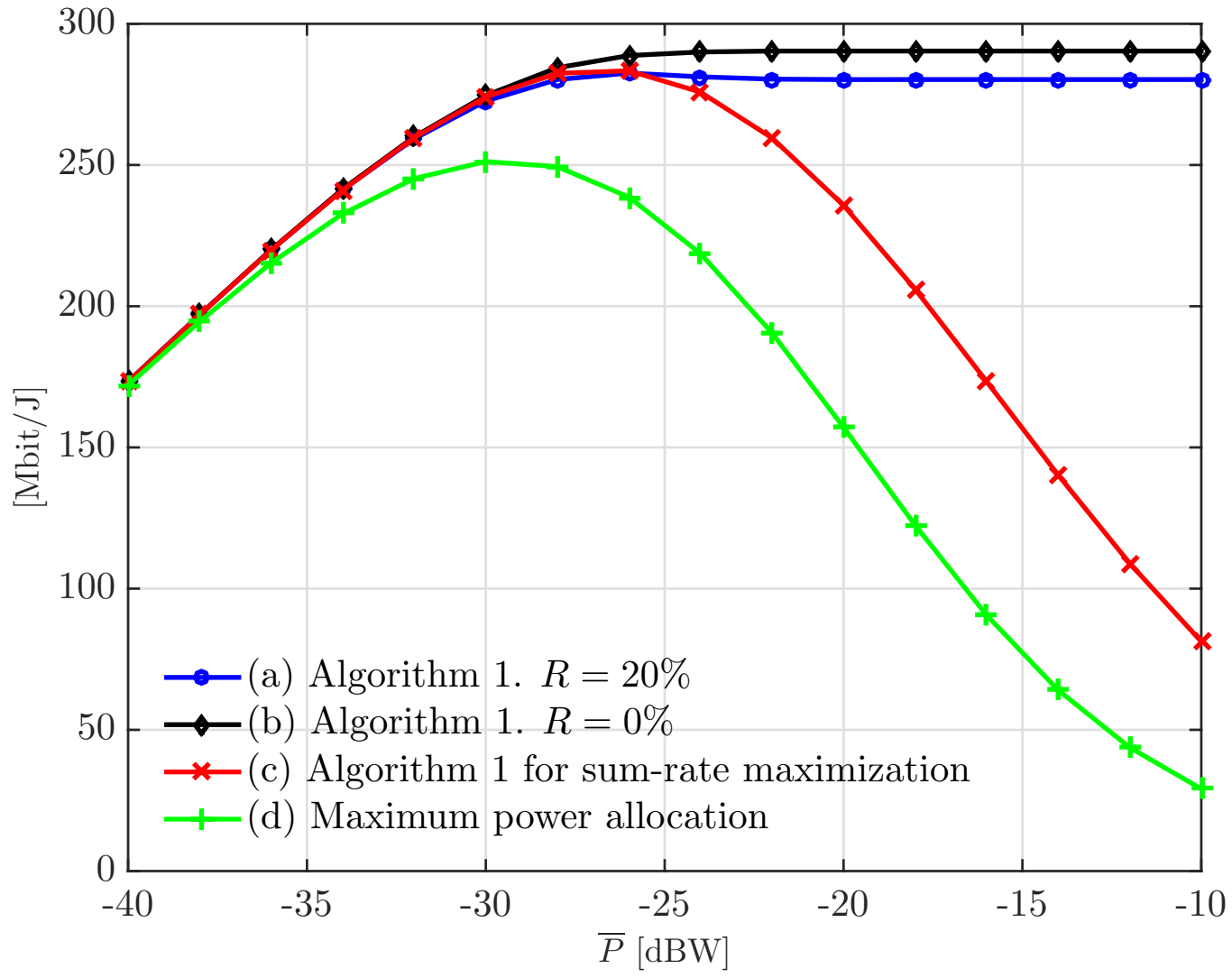

Figure 2. $K=5; M=50; \tau=10^{-2}$. Average achieved GEE versus $\overline{P}$ for: (a) Algorithm 1 for GEE maximization with $R=20\%$; (b) Algorithm 1 for GEE maximization with $R=0\%$; (c) sum-rate maximization by adapted Algorithm 1; (d) Maximum transmit power allocation.

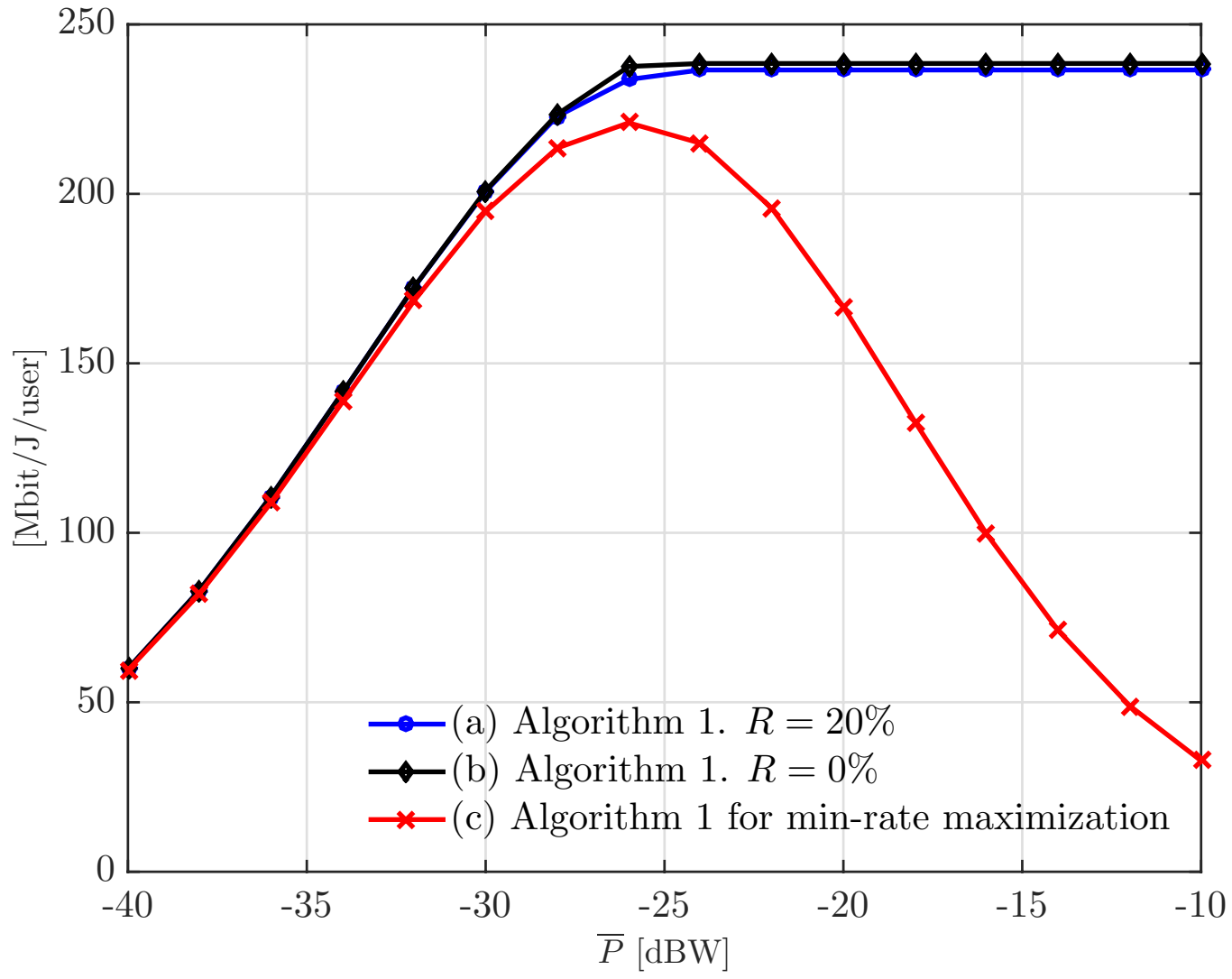

Figure 3. $K=5; M=50; \tau=10^{-2}$. Average achieved minimum users' energy efficiency versus $\overline{P}$ for: (a) Algorithm 1 for Min-EE maximization with $R=20\%$; (b) Algorithm 1 for Min-EE maximization with $R=0\%$; (c) Min-rate maximization by adapted Algorithm 1.

$R=0\%$; (c) Algorithm 1 specialized to maximize the sum-rate; (d) Maximum power allocation, i.e. $p_k=\overline{P}$ for all $k$, considered as a baseline scheme. In scheme (a), if the problem turns out to be unfeasible, the QoS constraint is relaxed and the solution from scheme (b) is taken. For low values of $\overline{P}$, this circumstance is very frequent, and indeed schemes (a) and (b) perform similarly. At the same time, schemes (b) and (c) also perform similarly for low $\overline{P}$, thus suggesting that in this range of $\overline{P}$, GEE and sum-rate maximization are equivalent. Instead, different performance are achieved for larger values of $\overline{P}$. Indeed, increasing $\overline{P}$ eventually allows attaining the peak of the GEE. At this point, the GEE achieved by scheme (b) remains constant, as using the excess transmit power would only decrease the GEE. Indeed, the GEE achieved by scheme (c) decreases, because this scheme makes use of the excess transmit power to maximize the sum-rate. Instead, scheme (a) strikes a balance between these two extremes. Some of the excess power is used to fulfill the QoS constraints, which results in a slightly lower GEE. However, once the constraints are met, the transmit power is not further increased and the achieved GEE keeps constant. We remark that this slight reduction of the GEE grants a higher minimum rate. In particular, in the saturation region of the GEE, the average minimum rate granted by scheme (b) is 1.6 [bit/s/Hz/user], whereas it increases to 2.35 [bit/s/Hz/user] when scheme (a) is used.

Similar considerations can also be made when Algorithm 1 is used to maximize the minimum of the users' EEs. Fig. 3 compares the minimum EE (with $w_k=1$ for all $k$) versus $\overline{P}$, achieved by: (a) Algorithm 1 with $R=20\%$; (b) Algorithm 1 with $R=0\%$; (c) min-rate maximization by adapted Algorithm 1. The results show a similar behavior as for Fig. 2.

Fig. 4 compares the GEE performance of the centralized Algorithm 1 and of its distributed counterpart Algorithm 2 for $R=0\%$ and $20\%$. We observe that, while the centralized scheme suffers a little performance gap in terms of GEE when QoS constraints are introduced, having minimum rate requirements causes a larger GEE degradation in the distributed scenario, especially for increasing $\overline{P}$. This is expected because unlike the centralized scheme, in the distributed setting the interference among the users is not jointly managed, which results in high multi-user interference, especially for large $\overline{P}$.

Table II shows the average number of iterations required by Algorithms 1 and 2 to converge when $R=0\%$ and $R=20\%$. Convergence is declared when $\|\mathbf{q}^{(i)}-\mathbf{q}^{(i-1)}\|^2/\|\mathbf{q}^{(i)}\|^2\leq 10^{-4}$. It is seen that both algorithms converge after a small number of iterations, which slightly increases for larger values

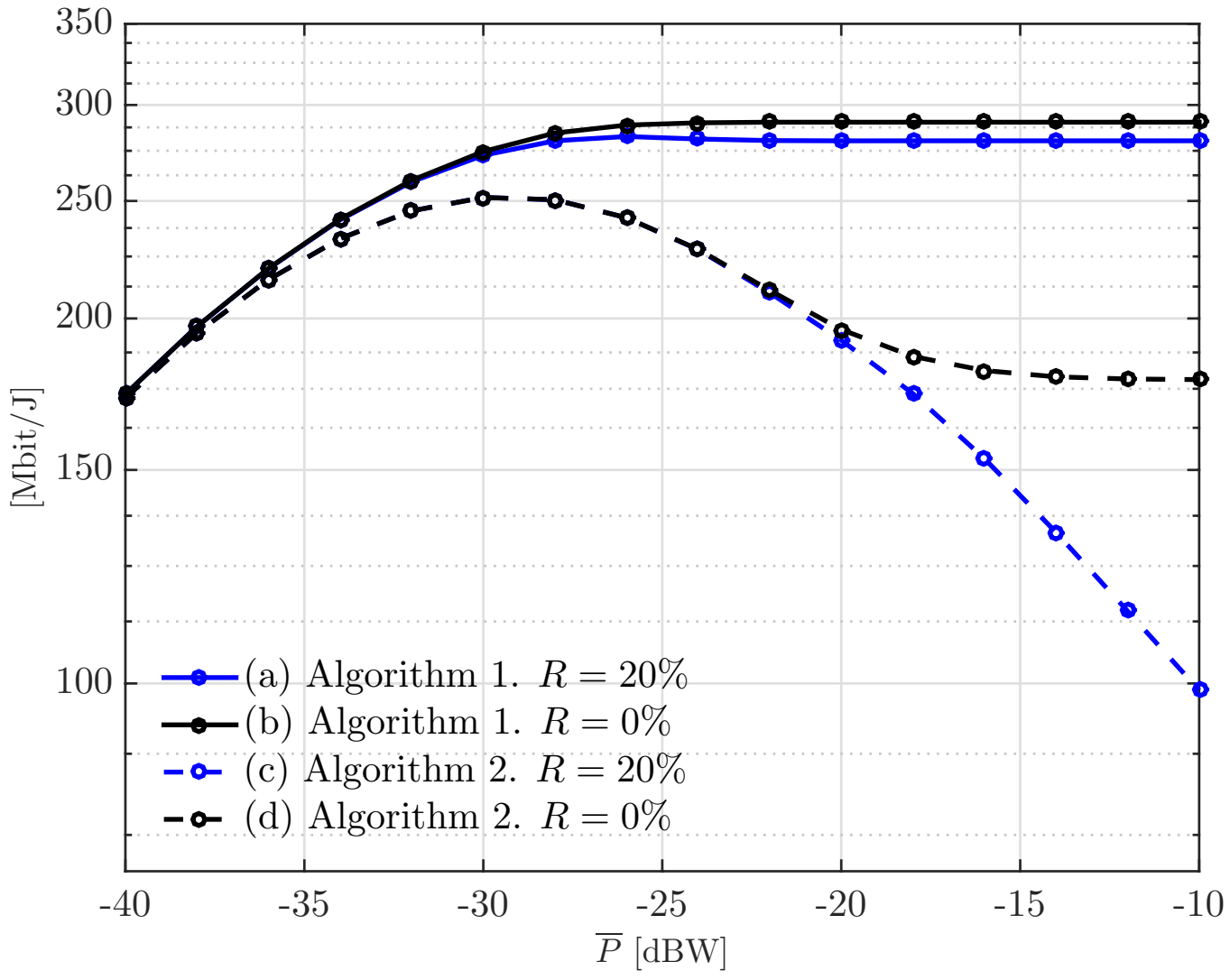

Figure 4. $K = 5; M = 50; \tau = 10^{-2}$. Average achieved GEE versus $\overline{P}$ for: (a) Algorithm 1 for GEE maximization with $R = 20\%$; (b) Algorithm 1 for GEE maximization with $R = 0\%$; (c) Algorithm 2 for distributed resource allocation with $R = 20\%$; (d) Algorithm 2 for distributed resource allocation with $R = 0\%$.

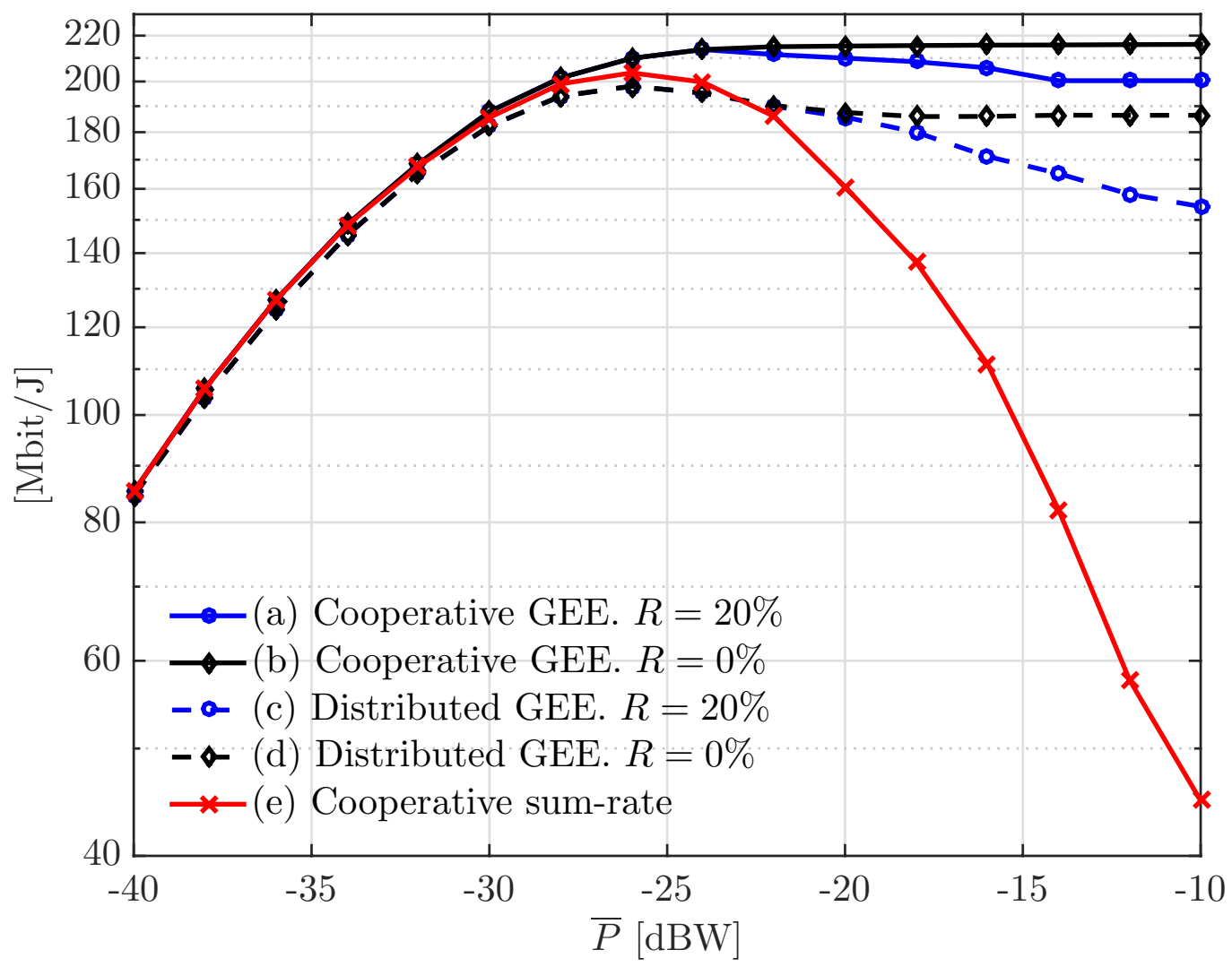

Figure 5. $K = 3; N = 16, M = 3$. Average achieved GEE versus $\overline{P}$ for: (a) Algorithm 3 for GEE maximization with $R = 20\%$; (b) Algorithm 3 for GEE maximization with $R = 0\%$; (c) Algorithm 4 for distributed resource allocation with $R = 20\%$; (d) Algorithm 4 for distributed resource allocation with $R = 0\%$; (e) sum-rate maximization by adapted Algorithm 3

of $\overline{P}$. This is because increasing $\overline{P}$ results in a larger feasible set. Observe that the distributed algorithm exhibits faster convergence than the centralized one. This makes it particularly suitable for self-organizing networks.

## B. *Relay-assisted OFDMA interference network*

Consider a relay-assisted multi-cell network as described in Section II-C2. Assume $S = 3$, $N = 16$, $K = 3$, $M = 3$, and $B = 180\,\text{kHz}$. The UEs are placed at a distance from the relay randomly generated in the interval $[100; 300]$ m and the same path-loss model and noise parameters as for previous figures are used.

Fig. 5 reports the GEE versus $\overline{P}$ achieved by the following resource allocation algorithms: a) Algorithm 3 with $R = 20\%$; b) Algorithm 3 with $R = 0\%$; c) Algorithm 4 with $R = 20\%$; d) Algorithm 4 with $R = 0\%$; e) Algorithm 3 specialized to maximize the sum-rate. If the problems with QoS constraints happen to be infeasible, the solution of the corresponding scheme with $R = 0\%$ is taken. Similar remarks can be made as for the massive MIMO system. In particular, by introducing QoS constraints we trade-off a slight reduction in GEE with a significant increase of the users' minimum rate, which, for the simulated scenario, increases from 3.8 [bit/s/Hz/user] when $R = 0\%$, to 7.16 [bit/s/Hz/user] when $R = 20\%$. As for the comparison between the centralized and decentralized approaches, we observe that the gap is rather limited in this scenario. This might be due to both the channel diversity granted by the use of multiple sub-carriers and to the rather limited number of non-cooperating transmitters. Indeed, it is known that that a limited number of players tend to result in a lower price-of-anarchy in game-theoretic settings [45].

## C. *Computational complexity discussion*

The computational complexity of the proposed algorithms depends on the number of iterations required to reach convergence as well as on the complexity of each iteration. A theoretical analysis for the number of iterations to reach convergence is very challenging. In fact, no theoretical results are available also for simpler scenarios than energy efficiency maximization. To overcome this problem, we have resorted to a numerical analysis. From Table II, it follows that the number of outer iterations for the single-resource-block case is very limited for both centralized and distributed algorithms (the latter require a slightly lower number of iterations). Similar results are obtained also in the multi-resource-block setting. The complexity of each iteration is analyzed in the following. The centralized algorithms are considered first.

*1) Centralized Algorithms:* We begin by observing that the complexity of the centralized algorithms depends on the complexity of each fractional problem to solve. To fix ideas, let us consider the maximization of the GEE – similar considerations apply to the maximization of the Min-EE. Although there exist several different methods to solve fractional programming problems, the most widely used is the Dinkelbach's algorithm. As shown in Algorithm 5, Dinkelbach's algorithm works by finding the solution of an auxiliary problem in each iteration, say $\mathbf{x}_n^*$, and then updating the parameter $\lambda$ as $\lambda_n = f(\mathbf{x}_n^*)/g(\mathbf{x}_n^*)$. One well-known result of the Dinkelbach's algorithm is that the sequence $\{\lambda_n\}_n$ converges with a super-linear rate. Moreover, this convergence rate does not depend on the complexity required to compute $\mathbf{x}_n^*$. In this respect, observe also that the auxiliary problems to obtain $\{\mathbf{x}_n^*\}$ are convex and therefore can be solved with polynomial complexity in the number of variables and constraints (which are $KN$ and $2K$, respectively). Putting all these facts together, we can conclude that the centralized algorithms have a polynomial complexity in each stage. This makes them of affordable complexity.

*2) Decentralized Algorithms:* As for the distributed case, a similar analysis can be performed since the proposed algorithms consist of an outer loop, which requires to solve at each stage $K$ fractional problems with $N$ variables and 2 constraints. Following the above arguments, we can state that the algorithms have a polynomial complexity in each stage. Observe that the complexity of each stage does not depend on $K$ (as in the centralized case) but only on the number of resource blocks $N$ and on the number of constraints (which is two). The parameter $K$ only scales the complexity of each iteration in a multiplicative way.

## VIII. Conclusion

In this paper, we have proposed a framework to develop centralized and decentralized power control algorithms for EE optimization in wireless networks. Unlike most previous related works, we have considered rate constraints and a more general SINR expression so as to encompass emerging 5G technologies. The resulting optimization problems have been tackled by an interplay of fractional programming, sequential convex optimization, and game theory. This has allowed us to derive centralized algorithms achieving first-order optimal points of the GEE and of the minimum of the users' EEs, and to develop decentralized algorithms that have been shown to converge to the GNE of the associated game. The analysis above has been performed in the case in which a single or multiple resource blocks are used for transmission. Numerical results have been used to corroborate the theoretical results. For this purpose, two case-studies have been considered: a hardware-impaired massive MIMO network and a multi-cell multi-carrier relay-assisted network.

The numerical analysis indicates that the centralized algorithms are quite robust to the enforcement of demanding rate constraints. Instead, the distributed algorithms are more sensitive to rate constraints, especially for increasing maximum feasible powers, due to the lack of centralized interference management. Also, the centralized algorithms perform better than their distributed counterparts, both with and without rate constraints, at the expense of a higher computational complexity and feedback requirements.

An important follow-up of this work is to analyze the gap between the proposed centralized framework and the global optimum of the GEE and minimum EE. To this end, an optimization framework is required, which allows one to effectively determine the global solution of energy-efficient optimization problems in interference-limited networks. This is a challenging problem, which is still much open. An attempt in this direction is taken in [46], wherein tools borrowed from monotonic optimization theory, fractional programming, and sequential optimization are combined together to characterize the global maximum of the energy efficiency in interference-limited wireless networks.

## Appendix A

Major concepts from fractional programming are reviewed here. For a more comprehensive overview, we refer to [16].

**Definition 1** (Fractional program). *Let $\mathcal{C} \subseteq \mathbb{R}^n$ be a convex set, and consider the functions $f : \mathcal{C} \to \mathbb{R}$ and $g : \mathcal{C} \to \mathbb{R}^+$. A fractional program is the optimization problem*

$$\max_{\mathbf{x} \in \mathcal{C}} \frac{f(\mathbf{x})}{g(\mathbf{x})}. \tag{88}$$

**Proposition 8** ([39]). *An $\mathbf{x}^* \in \mathcal{C}$ solves (88) if and only if $\mathbf{x}^* = \arg\max_{\mathbf{x}\in\mathcal{C}} \{f(\mathbf{x}) - \lambda^* g(\mathbf{x})\}$, with $\lambda^*$ being the unique zero of $F(\lambda) = \max_{\mathbf{x}\in\mathcal{C}} \{f(\mathbf{x}) - \lambda g(\mathbf{x})\}$.*

This result allows us to solve (88) by finding the zero of $F(\lambda)$. An efficient algorithm to do so is the Dinkelbach's algorithm [39], reported in Algorithm 5 for the reader's convenience. If $f(\mathbf{x})$ and $g(\mathbf{x})$ are concave and convex, respectively, then the Dinkelbach's algorithm requires to solve one convex problem in each iteration.[9] Moreover, the convergence rate of Dinkelbach's algorithm is known to be super-linear [39].

**Algorithm 5** Dinkelbach's algorithm

Set $\varepsilon > 0$; $\lambda = 0$;
**repeat**
  $\mathbf{x}^* = \arg\max_{\mathbf{x}\in\mathcal{C}} \{f(\mathbf{x}) - \lambda g(\mathbf{x})\}$
  $F = f(\mathbf{x}^*) - \lambda g(\mathbf{x}^*)$;
  $\lambda = f(\mathbf{x}^*)/ g(\mathbf{x}^*)$;
**until** $F \leq \varepsilon$

A considerable extension of (88) is to consider the maximization of the minimum of a set of ratios $\{f_i(\mathbf{x})/g_i(\mathbf{x})\}_{i=1}^I$. This problem is usually referred to as *generalized* fractional programming, and has been first studied in [40], wherein an optimization procedure is provided, based on a modification of Dinkelbach's algorithm. Specifically, the auxiliary function to be considered is $F(\lambda) = \min_{1\leq i\leq I} \{f_i(\mathbf{x}) - \lambda g_i(\mathbf{x})\}$, and the algorithm works as shown in Algorithm 6.

**Algorithm 6** Generalized Dinkelbach's algorithm

Set $\varepsilon > 0$; $\lambda = 0$;
**repeat**
  $\mathbf{x}^* = \arg\max_{\mathbf{x}\in\mathcal{C}} \{\min_{1\leq i\leq I} [f_i(\mathbf{x}) - \lambda g_i(\mathbf{x})]\}$;
  $F = \min_{1\leq i\leq I} \{f_i(\mathbf{x}^*) - \lambda g_i(\mathbf{x}^*)\}$;
  $\lambda = \min_{1\leq i\leq I} f_i(\mathbf{x}^*)/ g_i(\mathbf{x}^*)$;
**until** $F < \varepsilon$

Then, if each ratio has a concave numerator and a convex denominator, we can solve the generalized fractional problem by solving a sequence of convex problems,[10] with a linear convergence rate [40].

## Appendix B

In the $i$-th iteration of Algorithm 1, we solve (41) and compute the corresponding vector $\mathbf{p}^{(i)}$ of transmit powers, which maximizes the lower-bound $\tilde{\psi}_i$ at the $i$-th iteration, as given by (39), subject to the same constraints of the original problem (28). Then, the following chain of inequalities holds

$$\psi(\mathbf{p}^{(i)}) \geq \tilde{\psi}_i(\mathbf{p}^{(i)}) \geq \tilde{\psi}_i(\mathbf{p}^{(i-1)}) = \psi(\mathbf{p}^{(i-1)}) \tag{89}$$

[9] It is also required that $\lambda \geq 0$ in each iteration. This can be shown to always hold if the algorithm starts with $\lambda = 0$, and provided $\max_{\boldsymbol{x}} f(\boldsymbol{x}) \geq 0$.

[10] Observe that the minimum of concave functions is also concave [38].

wherein the first inequality follows because $\tilde{\psi}_i$ is a lower-bound of $\psi$, the second inequality follows because $\mathbf{p}^{(i)}$ is the maximizer of $\tilde{\psi}_i$, while the final equality holds because the parameters $a_k^{(i)}$ and $b_k^{(i)}$ in $\tilde{\psi}_i$ are such that the bound is tight in $\mathbf{p}^{(i-1)}$. As a consequence of (89), the value of $\psi$ increases after each iteration, and the algorithm must converge because $\psi$ is upper-bounded.

Next, let us denote by $\tilde{\psi}$ and $\tilde{\mathbf{p}}$ the lower-bound of $\psi$ and the power vector at convergence. By construction, $\tilde{\mathbf{p}}$ maximizes $\tilde{\psi}$ subject to the same constraints of the original problem (28), and therefore fulfills the associated KKT conditions. Such KKT conditions are the same as the KKT conditions of the original problem (28), except for the different objective function. However, upon convergence of Algorithm 1, we have $\tilde{\psi}(\tilde{\mathbf{p}}) = \psi(\tilde{\mathbf{p}})$ and $\nabla\tilde{\psi}(\tilde{\mathbf{p}}) = \nabla\psi(\tilde{\mathbf{p}})$, and the thesis follows.

## Appendix C

The first part of the proof follows along the same lines of Proposition 1. As for the KKT conditions, we remark that in this case we can not directly consider the KKT conditions of (29), because the objective of (29) is not differentiable. We can remove the non-differentiability by expressing (29) in its equivalent epigraph form:

$$\max_{\{t\in\mathbb{R}_+,\mathbf{p}\in\mathcal{P}\}} t \quad \text{subject to } \eta_k \geq t \quad \forall k. \tag{90}$$

Now, let us consider a modified version of Algorithm 1, which operates on (90) rather than on (29). In each iteration of this modified Algorithm 1, (37) is again used to lower-bound $\eta_k$ with $\tilde{\eta}_k$. This amounts to solving an approximation of (90), with $\tilde{\eta}_k$ in place of $\eta_k$, and updating the parameters $a_k^{(i)}$ and $b_k^{(i)}$ as in Algorithm 1. Following similar arguments as in the proof of Proposition 1, it follows that this modified version of Algorithm 1 converges to a point fulfilling the KKT conditions of (90), the only difference being that now the lower-bound is computed for the constraint function $\eta_k$ rather than for the objective function. However, since $\eta_k \geq \tilde{\eta}_k$, the solution of the approximate problem is always in the feasible set of (90). Moreover, upon convergence, $\eta_k$ and $\tilde{\eta}_k$ are equal, and so are their gradients. Therefore, the modified version of Algorithm 1 yields a power vector fulfilling the KKT conditions of (90). Observe now that replacing $\eta_k$ with $\tilde{\eta}_k$ in (90) yields the epigraph form of (45), i.e., the problem which is solved in each iteration of Algorithm 1 by the Generalized Dinkelbach's method. Hence, Algorithm 1 and its modified version (considered in this proof) converge to the same power allocation vector $\tilde{\mathbf{p}}$, and hence the thesis.

## Appendix D

Using [8, Sect. II.A], it follows that solving the EE problem (47) is equivalent to finding the root of the nonlinear function

$$\Phi(\lambda_k) = \max_{p_k\in\mathbb{R}_+} \{B\log(1+\gamma_k) - \lambda_k(p_{c,k}+p_k)\} \tag{91}$$

where $\lambda_k^\star \in \mathbb{R}_+$ is such that $\Phi(\lambda_k^\star) = 0$ and can be obtained through the Dinkelbach's algorithm. Setting to zero the derivative of (91) with respect to $p_k$ yields:

$$\frac{B}{1+\gamma_k}\frac{\partial\gamma_k}{\partial p_k} - \lambda_k = 0. \tag{92}$$

From (15), using (14) one gets $\partial\gamma_k/\partial p_k = \mu_k(1-\frac{\gamma_k}{\overline{\gamma}_k})^2$, from which (92) reduces to:

$$\frac{B\mu_k}{1+\gamma_k}\left(1-\frac{\gamma_k}{\overline{\gamma}_k}\right)^2 - \lambda_k = 0. \tag{93}$$

In the attempt of solving (91), let us study the properties of (93) as a function of $\gamma_k$. This amounts to studying $f_k(\gamma_k) = 0$ with $f_k(\gamma_k) = B\mu_k\gamma_k^2 - \left(2B\mu_k\overline{\gamma}_k + \lambda_k\overline{\gamma}_k^2\right)\gamma_k + \overline{\gamma}_k^2\left(B\mu_k - \lambda_k\right)$. The first derivative of $f_k(\gamma_k)$ with respect to $\gamma_k$ is

$$\frac{\partial f_k(\gamma_k)}{\partial\gamma_k} = 2B\mu_k\left(\gamma_k - \overline{\gamma}_k\right) - \lambda_k\overline{\gamma}_k^2 < 0 \tag{94}$$

where the last inequality follows since $0 \leq \gamma_k \leq \overline{\gamma}_k$, and $\mu_k \geq 0$, $\lambda_k > 0$, and $\overline{\gamma}_k > 0$. As a consequence, $f_k(\gamma_k)$ is a strictly decreasing function, which admits a solution if and only if $f_k(0) \geq 0$ and $f_k(\overline{\gamma}_k) \leq 0$ (with the equalities not simultaneously active). Since $f_k(\overline{\gamma}_k) = -\lambda_k\overline{\gamma}_k^2\left(1+\overline{\gamma}_k\right) < 0$, we need to ensure $f_k(0) = \overline{\gamma}_k^2\left(B\mu_k - \lambda_k\right) \geq 0$. This translates into $B\mu_k \geq \lambda_k$. By solving the second-order equation $f_k(\gamma_k) = 0$, one gets $\gamma_k = \nu_k\left(\lambda_k\right)$ where $\nu_k\left(x\right)$ is defined in (58). Plugging $\gamma_k = \nu_k\left(\lambda_k\right)$ into (14) yields (60).

## Appendix E

Rewrite $\mu_{k,n}$ in (11) as $\mu_{k,n} = \alpha_{k,n}/(\sigma_{k,n}^2 + I_{k,n})$ where $I_{k,n} = \sum_{j\neq k}\omega_{kj,n}p_{j,n}$. Using [17, Theorem 4], the GNE is unique if $\|\partial\mathbf{I}_k/\partial\mathbf{p}_{-k}\|\sup_{\mathbf{I}_k\in\mathbb{R}_+^N}\|\boldsymbol{\mathcal{B}}_k(\mathbf{p}_{-k})/\partial\mathbf{I}_k\| < 1$, for all $k$. with $\mathbf{I}_k = [I_{k,1},\ldots,I_{k,N}]^T$. From [17, Eq. (19)], we have

$$\|\partial\mathbf{I}_k/\partial\mathbf{p}_{-k}\| = \sqrt{\sum_{j=1,j\neq k}^{K}\sum_{n=1}^{N}\omega_{kj,n}^2} \tag{95}$$

$$\|\,\partial\boldsymbol{\mathcal{B}}_k(\mathbf{p}_{-k})/\,\partial\mathbf{I}_k\| = \sqrt{\sum_{\ell=1}^{N}\sum_{n=1}^{N}\left|\partial\pi_{k,n}(\lambda_k')\,/\partial I_{k,\ell}\right|^2}. \tag{96}$$

Observe now that $\pi_{k,n}(\lambda_k') > 0$ when $\mu_{k,n} > \lambda_k'$ (see Appendix D). After lengthy computations (not shown for space limitations), one gets

$$\frac{\partial p_{k,n}^\star}{\partial I_{k,\ell}} = \left[s_{k,n}\mathbb{1}_{n=\ell} + \frac{\overline{\gamma}_{k,n}}{g_{k,n}\left(\lambda_k'\right)}\xi_{k,\ell}\right]\mathbb{1}_{\mu_{k,n}>\lambda_k'} \tag{97}$$

where we have defined (for notational compactness) $s_{k,n}$ as in (83), whereas $\xi_{k,\ell}$ is defined in the text of Proposition 7. Plugging (97) into the RHS of (96) yields

$$\sqrt{\sum_{\ell\in\mathcal{S}_k^\star}\sum_{n\in\mathcal{S}_k^\star}\left[s_{k,n}\mathbb{1}_{n=\ell} + \frac{\overline{\gamma}_{k,n}}{g_{k,n}\left(\lambda_k'\right)}\xi_{k,\ell}\right]^2} \tag{98}$$

where $\mathcal{S}_k^\star \triangleq \{n=1,\ldots,N : \mu_{k,n} > \lambda_k'\}$. Putting all the above results together, Proposition 7 follows.

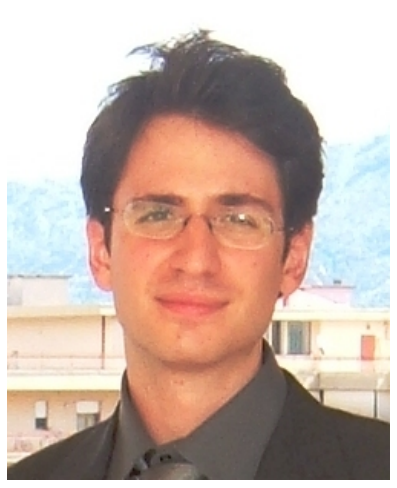

**Alessio Zappone** (S'08 - M'11) is a research associate at the Technische Universität Dresden, Dresden, Germany. Alessio received his M.Sc. and Ph.D. both from the University of Cassino and Southern Lazio. Afterwards, he worked with Consorzio Nazionale Interuniversitario per le Telecomunicazioni (CNIT) in the framework of the FP7 EU-funded project TREND, which focused on energy efficiency in communication networks. Since 2012, Alessio is the project leader of the project CEMRIN on energy-efficient resource allocation in wireless networks, funded by the German research foundation (DFG).

His research interests lie in the area of communication theory and signal processing, with main focus on optimization techniques for resource allocation and energy efficiency maximization. He held several research appointments at TU Dresden, Politecnico di Torino, Suplec - Alcatel-Lucent Chair on Flexible Radio, and University of Naples Federico II. He was the recipient of a Newcom# mobility grant in 2014. Alessio currently serves as associate editor for the IEEE SIGNAL PROCESSING LETTERS and IEEE JOURNAL ON SELECTED AREAS ON COMMUNICATIONS (Special Issue on Energy-Efficient Techniques for 5G Wireless Communication Systems).

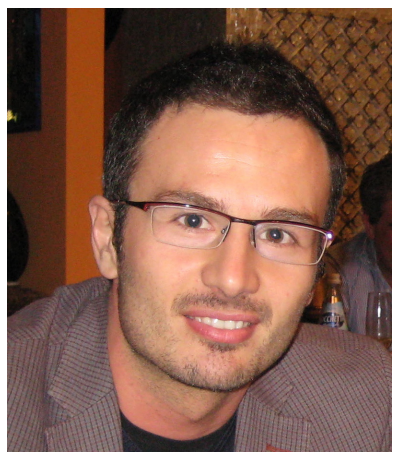

**Luca Sanguinetti** (SM'15) received the Laurea Telecommunications Engineer degree (cum laude) and the Ph.D. degree in information engineering from the University of Pisa, Italy, in 2002 and 2005, respectively. Since 2005 he has been with the Dipartimento di Ingegneria dell'Informazione of the University of Pisa. In 2004, he was a visiting Ph.D. student at the German Aerospace Center (DLR), Oberpfaffenhofen, Germany. During the period June 2007 - June 2008, he was a postdoctoral associate in the Dept. Electrical Engineering at Princeton. During the period June 2010 - Sept. 2010, he was selected for a research assistantship at the Technische Universität Munchen. From July 2013 to July 2015 he was with the Alcatel-Lucent Chair on Flexible Radio, Supélec, Gif-sur-Yvette, France. He is an Assistant Professor at the Dipartimento di Ingegneria dell'Informazione of the University of Pisa.

L. Sanguinetti is currently serving as an Associate Editor for the IEEE TRANSACTIONS ON WIRELESS COMMUNICATIONS, IEEE SIGNAL PROCESSING LETTERS and IEEE JOURNAL ON SELECTED AREAS OF COMMUNICATIONS (series on Green Communications and Networking). His expertise and general interests span the areas of communications and signal processing, game theory and random matrix theory for wireless communications. He was the co-recipient of two best paper awards: *IEEE Wireless Commun. and Networking Conference (WCNC) 2013* and *IEEE Wireless Commun. and Networking Conference (WCNC) 2014*. He is also the recipient of the FP7 Marie Curie IEF 2013 "Dense deployments for green cellular networks".

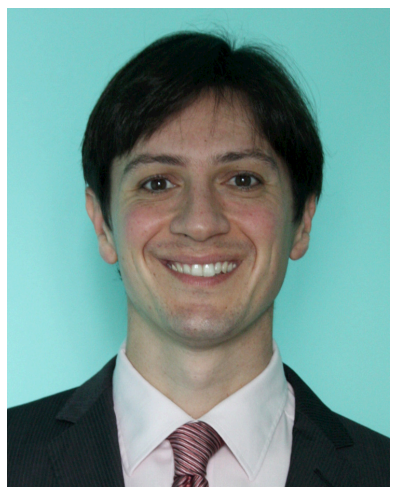

**Giacomo Bacci** (S'07–M'09) received the B.E. and the M.E. degrees in telecommunications engineering and the Ph.D. degree in information engineering from the University of Pisa, Pisa, Italy, in 2002, 2004, and 2008, respectively. In 2006-07, he was a visiting student research collaborator at Princeton University, Princeton, NJ, USA. In 2008-14, he was a post-doctoral research fellow at the University of Pisa. In 2008-2012, he also joined Wiser srl, Livorno, Italy, as a software engineer, and in 2012-14 he was also enrolled as a visiting postdoctoral research associate at Princeton University. Since 2015, he has joined MBI srl, Pisa, Italy, as a product manager for interactive satellite broadband communications.

Dr. Bacci is the recipient of the FP7 Marie Curie International Outgoing Fellowships for career development (IOF) 2011 GRAND-CRU, and received the Best Paper Award from the *IEEE Wireless Commun. and Networking Conference (WCNC) 2013*, the Best Student Paper Award from the *Intl. Waveform Diversity and Design Conf. (WDD) 2007*, and the Best Session Paper at the *ESA Workshop on EGNOS Performance and Applications 2005*. He is also the recipient of the 2014 URSI Young Scientist Award, and he has been named an Exemplary Reviewer 2012 for *IEEE Wireless Communications Letters*.

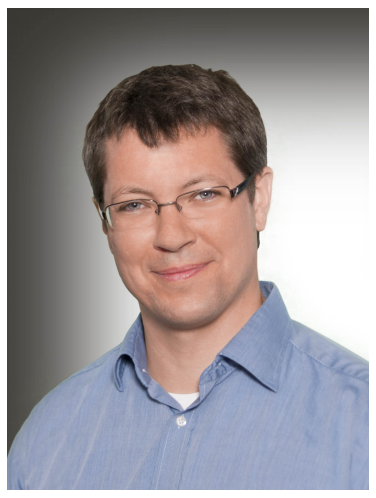

**Eduard Jorswieck** (S'01 – M'03 – SM'08) received the Diplom-Ingenieur (M.S.) degree and Doktor-Ingenieur (Ph.D.) degree, both in electrical engineering and computer science, from the Technische Universität Berlin, Germany, in 2000 and 2004, respectively. He was with the Broadband Mobile Communication Networks Department, Fraunhofer Institute for Telecommunications, Heinrich-Hertz-Institut, Berlin, from 2000 to 2008. From 2005 to 2008, he was a Lecturer with the Technische Universität Berlin. From 2006 to 2008, he was with the Department of Signals, Sensors and Systems, Royal Institute of Technology, as a Post-Doctoral Researcher and an Assistant Professor. Since 2008, he has been the Head of the Chair of Communications Theory and a Full Professor with the Technische Universität Dresden, Germany. He is principal investigator in the excellence cluster center for Advancing Electronics Dresden (cfAED) and founding member of the 5G lab Germany (5Glab.de).

His main research interests are in the area of signal processing for communications and networks, applied information theory, and communications theory. He has authored over 80 journal papers, 8 book chapters, some 225 conference papers and 3 monographs on these research topics. Eduard was a co-recipient of the IEEE Signal Processing Society Best Paper Award in 2006 and co-authored papers that won the Best Paper or Best Student Paper Awards at IEEE WPMC 2002, Chinacom 2010, IEEE CAMSAP 2011, IEEE SPAWC 2012, and IEEE WCSP 2012.

Dr. Jorswieck was a member of the IEEE SPCOM Technical Committee (2008 - 2013), and has been a member of the IEEE SAM Technical Committee since 2015. Since 2011, he has been an Associate Editor of the IEEE TRANSACTIONS ON SIGNAL PROCESSING. Since 2008, continuing until 2011, he has served as an Associate Editor of the IEEE SIGNAL PROCESSING LETTERS, and until 2013, as a Senior Associate Editor. Since 2013, he has served as an Editor of the IEEE TRANSACTIONS ON WIRELESS COMMUNICATIONS.

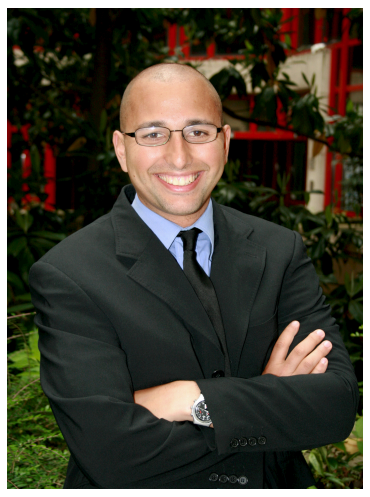

**Mérouane Debbah** (F'15) entered the Ecole Normale Suprieure de Cachan (France) in 1996 where he received his M.Sc and Ph.D. degrees respectively. He worked for Motorola Labs (Saclay, France) from 1999-2002 and the Vienna Research Center for Telecommunications (Vienna, Austria) until 2003. From 2003 to 2007, he joined the Mobile Communications department of the Institut Eurecom (Sophia Antipolis, France) as an Assistant Professor. Since 2007, he is a Full Professor at Supelec (Gif-sur-Yvette, France). From 2007 to 2014, he was director of the Alcatel-Lucent Chair on Flexible Radio. Since 2014, he is Vice-President of the Huawei France R&D center and director of the Mathematical and Algorithmic Sciences Lab. His research interests are in information theory, signal processing and wireless communications. He is an Associate Editor in Chief of the JOURNAL RANDOM MATRIX: THEORY AND APPLICATIONS and was an associate and senior area editor for IEEE TRANSACTIONS ON SIGNAL PROCESSING respectively in 2011-2013 and 2013-2014. Mérouane Debbah is a recipient of the ERC grant MORE (Advanced Mathematical Tools for Complex Network Engineering). He is a WWRF fellow and a member of the academic senate of Paris-Saclay. He is the recipient of the Mario Boella award in 2005, the 2007 IEEE GLOBECOM best paper award, the Wi-Opt 2009 best paper award, the 2010 Newcom++ best paper award, the WUN CogCom Best Paper 2012 and 2013 Award, the 2014 WCNC best paper award as well as the Valuetools 2007, Valuetools 2008, CrownCom2009 , Valuetools 2012 and SAM 2014 best student paper awards. In 2011, he received the IEEE Glavieux Prize Award and in 2012, the Qualcomm Innovation Prize Award.